\documentclass[a4paper,11pt]{article}
\pdfoutput=1 % if your are submitting a pdflatex (i.e. if you have
\usepackage{jheppub} % for details on the use of the package, please
\usepackage[T1]{fontenc} % if needed
\usepackage{multirow}
\usepackage{setspace}
\usepackage{makecell}
\usepackage{diagbox}

\title{Unitarity bounds on extensions of Higgs sector}

\author[a,1]{Bo-Qiang Lu}
\author[b,c,d,2]{Da Huang}

\affiliation[a]{School of Science, Huzhou University, Huzhou, Zhejiang 313000, China}
\affiliation[b]{National Astronomical Observatories, Chinese Academy of Sciences, Beijing, 100012, China}
\affiliation[c]{School of Fundamental Physics and Mathematical Sciences, Hangzhou Institute for Advanced Study, UCAS, Hangzhou 310024, China}
\affiliation[d]{International Centre for Theoretical Physics Asia-Pacific, Beijing/Hangzhou, China}

\emailAdd{bqlu@zjhu.edu.cn}
\emailAdd{dahuang@bao.ac.cn}

\abstract{
It is widely believed that extensions of the minimal Higgs sector is one of the promising directions for resolving many puzzles beyond 
the Standard Model (SM). 
In this work, we study the unitarity bounds on the models by extending the two-Higgs-doublet model with an additional real or complex Higgs 
triplet scalar. By noting that the SM gauge symmetries $SU(2)_L\times U(1)_Y$ are recovered at high energies, we can classify the two-body 
scattering states by decomposing the direct product of two scalar multiplets into their direct sum of irreducible representations of 
electroweak gauge groups. In such state bases, the s-wave amplitudes of two-body scalar scatterings can be written in the form of 
block-diagonalized scattering matrices. Then the application of the perturbative unitarity conditions on the eigenvalues of scattering matrices 
leads to the analytic constraints on the model parameters. {Finally, we numerically investigate the complex triplet scalar extension
of the two-Higgs-doublet model, finding that the perturbative unitarity places useful stringent bounds on the model parameter space.}%the trilinear scalar couplings and 
}

\begin{document} 
\maketitle
\flushbottom

\section{Introduction}

Despite the overwhelming success of the Standard Model (SM) by discovering the 125~GeV Higgs boson at the Large Hadron Collider (LHC)
in 2012~\cite{ATLAS:2012yve,CMS:2012qbp}, it has been widely believed that new physics is required to explain various phenomena beyond the SM, 
such as tiny neutrino masses~\cite{Formaggio:2021nfz}, the nature of dark matter~\cite{Planck:2015fie}, and the origin of matter-antimatter asymmetry in the Universe~\cite{Kuzmin:1985mm}.
One of the promising directions for resolving these puzzles is to extend the minimal SM Higgs sector by including additional scalars.
Note that the shape of the SM Higgs potential is fully determined by the vacuum expectation value (VEV), $v$, and the quartic self-coupling, $\lambda_H$. 
However, in the non-minimal extension of the Higgs section, there will be inevitable deviations of the Higgs self-couplings with respect to the SM predictions. 
Therefore, the precise measurement of the Higgs self-couplings can help us to probe the new physics and to understand the electroweak symmetry breaking mechanism. Until now, the determination of the trilinear Higgs coupling has been performed at the LHC Run 2 and will be further searched for at the Run 3, by directly detecting the single and double 
Higgs boson productions~\cite{Degrassi:2021uik} and other indirect probes~\cite{McCullough:2013rea,Cao:2015oxx,Bizon:2016wgr,deBlas:2016ojx}. In addition to the above experimental endeavors, 
%In addition to the experimental constraints on the Higgs sector from, for example, 
%the measurements of the trilinear Higgs coupling~\cite{Cao:2015oxx,Bizon:2016wgr} and the Higgs signal strengths~\cite{deBlas:2016ojx},
there has already been considerable theoretical explorations in order to constrain the Higgs sector, such as the perturbative unitarity~\cite{Gell-Mann:1969cuq,Weinberg:1971fb,Lee:1977yc,Lee:1977eg}, vacuum stability and triviality~\cite{Cabibbo:1979ay,Lindner:1985uk}. 

In this work, we shall focus on the systematic derivation of the perturbative unitarity bounds on the non-minimal Higgs sector with two or three Higgs multiplets.
As early as 1977, Lee, Quigg, and Thacker~\cite{Lee:1977yc,Lee:1977eg} made use of the perturbative unitarity and found the Higgs boson mass 
upper bound $m_{h}<870$~GeV in the minimal SM.
The perturbative unitarity has recently been calculated in various extensions of the Higgs sector and been identified as a significant constraint on the new physics.
One of the most popular extensions is the two-Higgs-doublet model (2HDM) (see Refs.~\cite{Branco:2011iw,Wang:2022yhm} for recent reviews) whose perturbative unitarity 
was firstly calculated in Refs.~\cite{Huffel:1980sk,Maalampi:1991fb,Kanemura:1993hm,Akeroyd:2000wc,Arhrib:2000is} 
with the assumptions of softly broken $Z_2$ symmetry and CP-conservation. %, i.e., real $m_{12}^2$ and $\lambda_5$ and $\lambda_6=\lambda_7=0$. 
The perturbative unitarity for the most general 2HDM was given in Ref.~\cite{Ginzburg:2005dt} and the associated numerical investigation% of the unitarity bound on the  general 2HDM 
was carried out in detail in Ref.~\cite{Kanemura:2015ska}.
For other Beyond-SM theories, %as far as we know, 
the unitarity bounds have been explored for the Georgi-Machacek model~\cite{Georgi:1985nv} in Ref.~\cite{Aoki:2007ah}, 
for the Type-II seesaw model~\cite{Konetschny:1977bn,Cheng:1980qt,Magg:1980ut,Schechter:1980gr,Lazarides:1980nt,Mohapatra:1980yp} 
in Ref.~\cite{Arhrib:2011uy}, for extended scalar sector with a real triplet scalar in Ref.~\cite{Khan:2016sxm}, 
for a complex triplet extension of the 2HDM with CP conservation and a softly broken $Z_2$ symmetry 
in Ref.~\cite{Ouazghour:2018mld}, and for the extension of SM with color-octet scalars in Ref.~\cite{Cao:2013wqa}, respectively. 
Some other applications of the unitarity bounds on new physics have been studied in Refs.~\cite{Chang:2019vez,Abu-Ajamieh:2021egq}.
% Refs.~\cite{Chang:2019vez,Abu-Ajamieh:2020yqi,Abu-Ajamieh:2021egq,Abu-Ajamieh:2022ppp}.}
%The unitarity bound for the extension with a complex Higgs triplet scalar (i.e., type-II seesaw model) or a real Higgs triplet scalar was calculated in Ref.~\cite{Arhrib:2011uy} and Ref.~\cite{Khan:2016sxm}, respectively. Ref.~\cite{Ouazghour:2018mld} calculated the perturbative unitarity for the extension of 2HDM with an additional complex Higgs triplet scalar, with the assumptions of softly broken $Z_2$ symmetry and CP-conservation.

In this paper, we systematically study the unitarity bounds in extensions of the 2HDM by including an additional real Higgs triplet $\Sigma$ 
with hypercharge $Y=0$ or a complex Higgs triplet scalar $\Delta$ with $Y=2$  in the most general setup, in order to ensure the validity of 
perturbation theory. Here we only concentrate on the high-energy limit where the SM gauge symmetry effects can be ignored.
%In order to ensure the perturbation of the theory, the scattering matrix is constrained by the unitarity limit. 
Thus, we can classify the two-scalar-particle states according to their conserved isospin and hypercharge quantum numbers, and construct the associated 2-to-2 scattering amplitude matrices in terms of the bases of $SU(2)_L \times U(1)_Y$ irreducible representations. %scattering processes according to the
%conserved quantum number, hypercharge and isospin, and construct the scattering matrices in terms of the bases of the irreducible representations.
Then we will consider the unitarity bounds in a few special cases, including the extensions of the SM or 2HDM by one additional real or complex scalar triplet, with or without a softly broken $Z_2$ symmetry. {Finally, we will numerically apply our derived perturbative unitarity bounds to the complex triplet extension of the 2HDM, and show the corresponding constraints on model parameter spaces.}
%study the application of the perturbative unitarity to various models and show the potential constraints on the Higgs scalars' mass.
%Using the bases provided in this work, one can also easily construct the scattering matrices for the models with three Higgs doublets and those with a Higgs doublet and a complex and a real triplet.

This work is organized as follows. In Sec.~\ref{sec:Bases}, we provide the two-particle state eigenbasis according to the irreducible 
representations of given hypercharges and isospins. In Secs.~\ref{sec:2HDMreal} and~\ref{sec:2HDMcomplex}, we present the scattering 
amplitude matrices in extensions of the 2HDM by including an additional real triplet and complex triplet, respectively. 
In Sec.~\ref{sec:application}, we show the constraints of unitarity bounds on model parameters in the extension of 2HDM with a complex triplet scalar. 
The conclusions are summarized in Sec.~\ref{sec:conclusions}. 
In Appendix~\ref{apd:mevs}, we provide the analytical solutions for a dimensional-five scattering matrix appearing in the real or complex 
triplet augment of the 2HDM with a softly broken $Z_2$ symmetry.
{The analytic relations between parameters in the generic scalar basis and the Higgs basis in the complex triplet extension of 2HDM are provided in Appendix~\ref{apd:HBcouplings}. 
In Appendix~\ref{apd:Mmatrix}, we provide elements of the scalar mass matrices in this model.}
Finally, the trilinear couplings of a neutral scalar with two charged Higgs particles are summarized in Appendix~\ref{apd:tricoupling}.

%%%%%%%%%%%%%%%%%%%%%%%%%%%%%%%%%%%%%%%%%%%%%%%%%%
\section{Two-particle eigenstates and unitarity bounds on  scattering matrices} 
\label{sec:Bases}
%%%%%%%%%%%%%%%%%%%%%%%%%%%%%%%%%%%%%%%%%%%%%%%%%%
The calculation of the unitarity bounds in the minimal SM was firstly investigated in Refs.~\cite{Lee:1977yc,Lee:1977eg} and has been applied to various extensions of the SM.
It requires that the eigenvalues of this scattering matrix should be less than the unitarity limit~\cite{Gell-Mann:1969cuq,Weinberg:1971fb,Ginzburg:2005dt}, 
otherwise the perturbative calculation of scattering amplitudes at tree level is no more reliable. % and the model may be unphysical.
From another perspective, one can make a partial wave expansion of the scattering amplitudes for the interaction channels and put the unitarity
bounds on the partial wave amplitudes.
Concretely, the cross section of scalar scattering processes $s_1s_2\to s_3s_4$ can be expressed in terms of the partial wave decomposition as
\begin{equation}
    \sigma=\frac{16 \pi}{s} \sum_{l=1}^{\infty}(2 l+1)\left|a_l(s)\right|^2,
\end{equation}
where $s$ is the Mandelstam variable and $a_l$ is the partial wave coefficients with the specific angular momenta $l$. Together with the optical theorem one finds
the following bound of unitarity: 
\begin{equation}\label{eq:UB}
    \left|\text{Re}\left(a_l\right)\right|<\frac{1}{2}\,, \quad \text{for all}~ l.
\end{equation}
In the high energy limit, it is found that the $s$-wave amplitude $a_0(s)$ is dominated by the point vertex processes 
since the $s$-, $t$-, $u$-channel processes are suppressed by the scattering energy. 
Furthermore, the equivalence theorem~\cite{Cornwall:1973tb,Cornwall:1974km,Yao:1988aj,Veltman:1989ud,He:1992nga} declares that at very high energy, 
the amplitudes of scattering processes involving longitudinal gauge bosons in the initial and final states are equivalent to those in which gauge bosons are replaced by the corresponding Nambu-Goldstone bosons. 
Thus, in the high energy limit $a_0(s)$ is fully determined by the quartic couplings of the scalar potential.

{Using the equivalence theorem, we can write down the two-particle state bases in terms of the components of the Higgs multiplets. 
Once given the scalar potential, we can determine the amplitudes for the $2\to 2$ scattering processes with the bases.
This largely simplifies the calculations for scattering amplitudes.}
In Refs.~\cite{Arhrib:2000is,Aoki:2007ah,Kanemura:2015ska,Ginzburg:2005dt}, the bases are further classified according to their electroweak (EW) charges, {\it i.e.}, total hypercharge $Y$ and total isospin $I$,
%The other way to classify the bases is based on their total hypercharge $Y$ and total isospin $I$, as proposed in Ref.~\cite{Ginzburg:2005dt}, 
since the EW $SU(2)_L \times U(1)_Y$ gauge symmetries are recovered at high energies so that their associated quantum numbers becomes conserved again. % quantum numbers in the scattering at high energy.
In this approach, we decompose the direct product of two Higgs multiplets into the direct sums of irreducible representations under EW gauge symmetries.% and does not require an explicit expansion of the Higgs multiplets.

\begin{table}[tbp]
    \renewcommand\arraystretch{1.5}
    \centering
    \begin{tabular}{|c|c|c|c|c|c|}\hline
        Field & $\Phi$ & $\tilde{\Phi}$ & $\Sigma$ & $\Delta$ & $\tilde{\Delta}$ \\ \hline
        $SU(2)_L$ isospin & 2 & 2 & 3 & 3 & 3 \\ \hline
        Hypercharge & 1 & $-1$ & 0 & 2 & $-2$ \\ \hline   
    \end{tabular}
    \caption{A summary of the quantum numbers of the Higgs multiplets. $\Phi$, $\Sigma$, and $\Delta$ denotes the $SU(2)_L$ doublet, real triplet, and complex triplet, 
    respectively. We define $\tilde{\Phi}=i \tau_2 \Phi^*$ and $\tilde{\Delta}_{ab}= (i \tau_2)_{ac} (i\tau_2)_{bd} (\Delta^\dagger)^{cd}$, which have negative hypercharge.} 
    \label{tab:fields}
\end{table}

In this work, we adopt an intermediate route for the classification of the bases. Firstly, we classify the direct product of the two
Higgs multiplets according to their total hypercharge, where the isospins and hypercharges of the Higgs multiplets considered in the present work are summarized in Table~\ref{tab:fields}. 
Then we decompose the direct product into direct sums of the irreducible representations of the EW $SU(2)_L$ symmetry. 
There are three types of direct products of Higgs multiplets we are concerned about, which are given as follows:
\begin{equation}
    2\otimes 2 = 1\oplus 3,~2\otimes 3 = 2\oplus 4,~{\rm and}~3\otimes 3 = 1\oplus 3\oplus 5.
\end{equation}
In this way, we %decompose the direct products and 
classify the two-particle bases according to their total isospins and hypercharges of the two Higgs multiplets. 
Furthermore, we express the bases of the irreducible representation in terms of %the un-physical 
components in the multiplets. 
The results are summarized in Tables~\ref{tab:dp22}-\ref{tab:dprealcomplex33}, in which the eigenstates are rescaled so that they are normalized. Moreover, due to the symmetry property when exchanging two identical bosons, some representations of the two-particle eigenstates vanish, {\it e.g.}, the $(Y,I)=(2,0)$ state in the $\Phi_i \times\Phi_j$ when $i=j$ in Table~\ref{tab:dp22}, the $(Y,I)=(0,1)$ state from $\Sigma\times\Sigma$ in Table~\ref{tab:dpreal33}, and the $(Y,I)=(4,0)$ state from $\Delta\times\Delta$ in Table~\ref{tab:dpcomplex33}.

\begin{table}[tbp]
    \renewcommand\arraystretch{1.8}
    \centering
    \begin{tabular}{|c|c|c|}\hline
        % \diagbox{Hypercharge}{Isospin} 
        & $I=0$ & $I=1$ \\ \hline
        $Y=0$ & $\frac{1}{\sqrt{2}}\left(w_i^+w_j^- + H_i^0H_j^{0*}\right)$ & 
        \makecell[c]{$ w_i^+H_j^{0*}$\\
        $\frac{1}{\sqrt{2}}\left( -w_i^+w_j^-+H_i^0H_j^{0*} \right)$\\
        $-H_i^0w_j^-$} \\ \hline
        $Y=2$ & $\frac{1}{\sqrt{2}}\left(-w_i^+H_j^0 + H_i^0w_j^{+}\right)$ & 
        \makecell[c]{$w_i^+w_j^+~\left(\times\frac{1}{\sqrt{2}}~{\rm for}~i=j\right)$\\
        $\frac{1}{\sqrt{2}}\left( w_i^+H_j^0 + H_i^0w_j^{+} \right)$\\
        $H_i^0H_j^0~\left(\times\frac{1}{\sqrt{2}}~{\rm for}~i=j\right)$}\\ \hline   
    \end{tabular} % }
    \caption{The bases of the irreducible representation for the two Higgs doublets direct product. The bases in the first and second row
    are corresponding to the direct product $\Phi_{i}\times \tilde{\Phi}_{j}$ $(Y=0)$ and $\Phi_{i}\times \Phi_{j}$ $(Y=2)$, respectively. 
    Note that $i$ and $j$ indicate the Higgs doublet. We observe that the bases with $(Y=2,~I=2)$ vanish when the two Higgs doublets are identical, i.e., $i=j$.}
    \label{tab:dp22}
\end{table}

\begin{table}[tbp]
    \renewcommand\arraystretch{1.8}
    \centering
    \begin{tabular}{|c|c|c|}\hline
         & $I=\frac{1}{2}$ & $I=\frac{3}{2}$ \\ \hline
        $Y=1$ & \makecell[c]{$\sqrt{\frac{2}{3}}\left(-\frac{i}{\sqrt{2}}w^+\sigma^0+H^0\sigma^+\right)$\\
        $\sqrt{\frac{2}{3}}\left(-w^+\sigma^-+\frac{i}{\sqrt{2}}H^0\sigma^0\right)$} & 
        \makecell[c]{$ w^+\sigma^+$\\
        $\frac{1}{\sqrt{3}}\left(i\sqrt{2} w^{+} \sigma^0+H^0 \sigma^{+}\right)$\\
        $\frac{1}{\sqrt{3}}\left(w^{+} \sigma^{-}+i\sqrt{2} H^0 \sigma^0\right)$\\
        $H^0 \sigma^{-}$} \\ \hline
    \end{tabular}
    \caption{The bases of the irreducible representation for the direct product of a Higgs doublet and a real Higgs triplet scalar, $\Phi \times \Sigma$.}
    \label{tab:dpreal23}
\end{table}

\begin{table}[tbp]
    \renewcommand\arraystretch{1.8}
    \centering
    \begin{tabular}{|c|c|c|}\hline
         & $I=\frac{1}{2}$ & $I=\frac{3}{2}$ \\ \hline
        $Y=1$ & \makecell[c]{$\sqrt{\frac{2}{3}}\left(-\frac{1}{\sqrt{2}} H^{0 *} \delta^{+}+w^{-} \delta^{++}\right)$\\
        $\sqrt{\frac{2}{3}}\left(-H^{0 *} \delta^0-\frac{1}{\sqrt{2}} w^{-} \delta^{+}\right)$} & 
        \makecell[c]{$-H^{0 *}\delta^{++}$\\
        $\frac{1}{\sqrt{3}}\left(\sqrt{2} H^{0 *} \delta^{+}+w^{-} \delta^{++}\right)$\\
        $\frac{1}{\sqrt{3}}\left(H^{0 *} \delta^0-\sqrt{2} w^{-} \delta^{+}\right)$\\
        $-w^{-} \delta^0$}\\ \hline
        $Y=3$ & \makecell[c]{$\sqrt{\frac{2}{3}}\left(-\frac{1}{\sqrt{2}} w^{+} \delta^{+}-H^0 \delta^{++}\right)$\\
        $\sqrt{\frac{2}{3}}\left(-w^{+} \delta^0+\frac{1}{\sqrt{2}} H^0 \delta^{+}\right)$} & 
        \makecell[c]{$-w^{+} \delta^{++}$\\
        $\frac{1}{\sqrt{3}}\left(\sqrt{2} w^{+} \delta^{+}-H^0 \delta^{++}\right)$\\
        $\frac{1}{\sqrt{3}}\left(w^{+} \delta^0+\sqrt{2} H^0 \delta^{+}\right)$\\
        $H^0 \delta^0$}\\ \hline   
    \end{tabular}
    \caption{The bases of the irreducible representation for the direct product of a Higgs doublet and a complex Higgs triplet scalar. 
    The bases in the first and second row are corresponding to the direct product $\tilde{\Phi}\times \Delta$ $(Y=1)$ and $\Phi\times \Delta$ $(Y=3)$, respectively.}
    \label{tab:dpcomplex23} 
\end{table}

\begin{table}[tbp]
    \renewcommand\arraystretch{1.8}
    \centering
    \begin{tabular}{|c|c|c|c|}\hline
         & $I=0$ & $I=1$ & $I=2$ \\ \hline
        $Y=0$ & $\sqrt{\frac{2}{3}}\left(\sigma^{+}\sigma^{-}+\frac{1}{2}\sigma^{0}\sigma^{0}\right)$ & 0 &
        \makecell[c]{$\frac{1}{\sqrt{2}}\sigma^+\sigma^+$\\
        $i\sigma^+\sigma^0$\\
        $\frac{1}{\sqrt{3}}\left( \sigma^+\sigma^--\sigma^0\sigma^0 \right)$\\
        $i\sigma^-\sigma^0$\\
        $\frac{1}{\sqrt{2}}\sigma^-\sigma^-$} \\ \hline
    \end{tabular}
    \caption{The bases of the irreducible representation for the two real Higgs triplets direct product, $\Sigma\times \Sigma$.
    We only consider the case of the direct product of two identical triplet scalar (i.e., $i=j$). We find the bases with $I=1$ vanish in this case.}
    \label{tab:dpreal33}
\end{table}

\begin{table}[tbp]
    \renewcommand\arraystretch{1.8}
    \centering
    \begin{tabular}{|c|c|c|c|}\hline
         & $I=0$ & $I=1$ & $I=2$ \\ \hline
        $Y=0$ & $\frac{1}{\sqrt{3}}\left(\delta^{++}\delta^{--}+\delta^{+}\delta^{-}+\delta^{0}\delta^{0*}\right)$ & 
        \makecell[c]{$\frac{1}{\sqrt{2}}\left( -\delta^{++}\delta^{-}+\delta^{+}\delta^{0*} \right)$\\
        $-\frac{1}{\sqrt{2}}\left( \delta^{++}\delta^{--}-\delta^{0}\delta^{0*} \right)$\\
        $-\frac{1}{\sqrt{2}}\left(-\delta^{+}\delta^{--}+\delta^{-}\delta^{0}\right)$ } &
        \makecell[c]{$-\delta^{++}\delta^{0*}$\\
        $\frac{1}{\sqrt{2}}\left( \delta^{++}\delta^{-}+\delta^{+}\delta^{0*}\right)$\\
        $\frac{1}{\sqrt{6}}\left( -2\delta^{+}\delta^{-}+\delta^{++}\delta^{--}+\delta^{0}\delta^{0*}\right)$\\
        $\frac{1}{\sqrt{2}}\left( -\delta^{+}\delta^{--}-\delta^{0}\delta^{-}\right)$\\
        $-\delta^{--}\delta^{0}$} \\ \hline
        $Y=4$ & $\sqrt{\frac{2}{3}}\left(-\delta^{++}\delta^{0}-\frac{1}{2}\delta^{+}\delta^{+}\right)$ & 0 & 
        \makecell[c]{$\frac{1}{\sqrt{2}}\delta^{++}\delta^{++}$\\
        $-\delta^{++}\delta^{+}$\\
        $\frac{1}{\sqrt{3}}\left( \delta^{+}\delta^{+}-\delta^{++}\delta^{0} \right)$\\
        $\delta^{0}\delta^{+}$\\
        $\frac{1}{\sqrt{2}}\delta^{0}\delta^{0}$} \\ \hline 
    \end{tabular}
    \caption{The bases of the irreducible representation for the direct product of two complex Higgs triplet scalars. The bases in the first and second row
    are corresponding to the direct product $\Delta\times \tilde{\Delta}$ $(Y=0)$ and $\Delta\times \Delta$ $(Y=4)$, respectively. 
    We observe again that the bases with $(Y=4,~I=1)$ vanish because the two triplets are identical.}
    \label{tab:dpcomplex33}
\end{table}

\begin{table}[tbp]
    \renewcommand\arraystretch{1.8}
    \centering
    \begin{tabular}{|c|c|c|c|}\hline
         & $I=0$ & $I=1$ & $I=2$ \\ \hline
        $Y=2$ & $\frac{1}{\sqrt{3}}\left(\sigma^{+}\delta^{0}-i\sigma^0\delta^{+}-\sigma^{-}\delta^{++}\right)$ & 
        \makecell[c]{$-\frac{1}{\sqrt{2}}\left( \sigma^+\delta^++i\sigma^0\delta^{++} \right)$\\
        $-\frac{1}{\sqrt{2}}\left( \sigma^+\delta^0+\sigma^-\delta^{++} \right)$\\
        $-\frac{1}{\sqrt{2}}\left( i\sigma^0\delta^0-\sigma^-\delta^{+} \right)$} &
        \makecell[c]{$\sigma^-\delta^{++}$\\
        $\frac{1}{\sqrt{2}}\left( \sigma^+\delta^{+}-i\sigma^0\delta^{++} \right)$\\
        $\frac{1}{\sqrt{6}}\left( \sigma^+\delta^{0}+i2\sigma^0\delta^{+}-\sigma^-\delta^{++} \right)$\\
        $\frac{1}{\sqrt{2}}\left( i\sigma^0\delta^{0}+\sigma^-\delta^{+} \right)$\\
        $\sigma^-\sigma^0$} \\ \hline
    \end{tabular}
    \caption{The bases of the irreducible representation for the direct product of a real Higgs triplet and a complex Higgs triplet, $\Sigma\times\Delta$.}
    \label{tab:dprealcomplex33}
\end{table}

Based on the above two-particle basis, we can determine the $2\to2$ scattering amplitudes as follows~\cite{Ginzburg:2005dt}
\begin{equation}
	S_{(Y, I)}=\left\langle(\phi \phi)_{Y,I}^f|\hat{S}|(\phi \phi)_{Y,I}^{i}\right\rangle,
\end{equation}
in the sector with definite EW charges $(Y,I)$ with $Y$ and $I$ as the total hypercharge and isospin, respectively. We do not distinguish states by the third components of isospin $I_3$, since the states with same $(Y,I)$ but different $I_3$ would lead to exactly the same scattering matrix. In the tree-level approximation,  the elements of the scattering matrix among scalars are determined by the quartic couplings in the scalar potential. Here we do not decompose a complex scalar field into its real and imaginary parts either in the external state basis or in the scalar potential due to the recovered $SU(2)_L\times U(1)_Y$ symmetry at high energies. 

As argued before, the unitarity of scattering amplitudes requires that the $s$-wave amplitude $a_0(s)$ in the partial-wave expansion should fulfill the bound in Eq.~(\ref{eq:UB}). Note that the amplitude of two scalar scatterings is dominated by the $s$-wave one at the tree level, so that the unitarity bounds can be transformed into the following condition on the eigenvalues $\Lambda_{(Y,I)}$ of the scattering matrices $16\pi S_{(Y,I)}$ as follows 
\begin{equation}\label{eq:UBonEV}
	|\Lambda_{(Y,I)}|\leq 8\pi\,.
\end{equation}

\section{Two-Higgs-doublet model plus a real triplet}\label{sec:2HDMreal}

In this section, we will focus on the model that contains two Higgs doublets and a real Higgs triplet scalar. The scattering matrix for the scalar potential is 
provided in sec.~\ref{sec:2hdmrealsc}. Using these results, we consider the perturbative unitarity constraints on two simplified cases: 
the model with a softly broken $Z_2$ symmetry and the $\Sigma$SM model~\cite{FileviezPerez:2008bj}. % (see the following section for more detail).}

\subsection{The scalar potential}

The scalar potential for the extension of the 2HDM with an additional real Higgs triplet field 
$\Sigma$ %under the $SU(2)_L\times U(1)_Y$ gauge symmetry 
is given by
\begin{equation}\label{eq:real2HDM}
    V_{r}=V\left(\Phi_1, \Phi_2\right)+V(\Sigma)+V\left(\Phi_1, \Phi_2, \Sigma\right),
    \end{equation}
where the Higgs doublet and real triplet scalar are
\begin{equation}\label{eq:hrcompont}
    \Phi_i =\begin{pmatrix}
        w_i^+\\ 
        H_i^0
    \end{pmatrix},~{\rm and}~
    \Sigma=\left(\begin{array}{cc}
    \sigma^{0} / \sqrt{2} & \sigma^{+} \\
    \sigma^{-} & -\sigma^{0} / \sqrt{2}
    \end{array}\right),
\end{equation}
where we can further expand the $H_i^0=\frac{1}{\sqrt{2}}(\varphi_i+iz_i)$.
The most general renormalizable scalar potential for the 2HDM in the generic basis $\{\Phi_1,\Phi_2\}$ is commonly written as~\cite{Davidson:2005cw}
\begin{equation}\label{eq:2HDM}
    \begin{aligned}
    V\left(\Phi_1, \Phi_2\right)=& m_1^2 \Phi_1^{\dagger} \Phi_1+m_2^2 \Phi_2^{\dagger} \Phi_2-\left(m_{12}^2\Phi_1^{\dagger} \Phi_2
    +\text {H.c.}\right)+\frac{1}{2} \lambda_1\left(\Phi_1^{\dagger} \Phi_1\right)^2 \\
    &+\frac{1}{2} \lambda_2\left(\Phi_2^{\dagger} \Phi_2\right)^2+\lambda_3 \Phi_1^{\dagger} \Phi_1 \Phi_2^{\dagger} \Phi_2
    +\lambda_4 \Phi_1^{\dagger} \Phi_2 \Phi_2^{\dagger} \Phi_1+\left[\frac{1}{2} \lambda_5\left(\Phi_1^{\dagger} \Phi_2\right)^2\right.\\
    &\left.+\lambda_6\left(\Phi_1^{\dagger} \Phi_1\right)\left(\Phi_2^{\dagger} \Phi_1\right)
    +\lambda_7\left(\Phi_2^{\dagger} \Phi_2\right)\left(\Phi_2^{\dagger} \Phi_1\right)+\text {H.c.}\right].
    \end{aligned}
\end{equation}
The parameters $m_{12}^2$, $\lambda_5$, $\lambda_6$, and $\lambda_{7}$ should be real if we impose the CP conservation on the potential.
If the $Z_2$ symmetry with $\Phi_1 \to \Phi_1$ and $\Phi_2 \to -\Phi_2$ is only softly broken by the term proportional to $m_{12}^2$, we should require $\lambda_6=\lambda_7=0$. 
The potential for the real Higgs triplet scalar is given by
\begin{equation}\label{VSigma}
    V(\Sigma)=\frac{1}{2}m_{\Sigma}^2 \operatorname{Tr}\Sigma^2+\frac{1}{4}\lambda_{\Sigma}\operatorname{Tr}\Sigma^4\,.
\end{equation}
while the interactions between the Higgs doublets and the real triplet read as follows 
\begin{eqnarray}\label{eq:intreal2hdmc}
    V(\Phi_1,\Phi_2,\Sigma)&=&\frac{1}{\sqrt{2}}\left[ a_1 \Phi_1^{\dagger} \Sigma \Phi_1+a_2 \Phi_2^{\dagger} 
    \Sigma \Phi_2 + \left( a_{12} \Phi_1^{\dagger} \Sigma \Phi_2 + \text {H.c.} \right ) \right]\nonumber\\
    &+&\frac{1}{2}\operatorname{Tr}\Sigma^2\left[\lambda_8 \Phi_1^{\dagger} \Phi_1+\lambda_9 \Phi_2^{\dagger} \Phi_2
    +\left(\lambda_{10} \Phi_1^{\dagger} \Phi_2+\text {H.c.}\right)\right]\,.
\end{eqnarray}
For a real triplet, the possible terms $\operatorname{Tr}\left(\Sigma^4\right)$ and $\Phi^{\dagger} \Sigma^2 \Phi$ are not independent since they can be expressed as the combination of $\left[\operatorname{Tr}\left(\Sigma^2\right)\right]^2$ and $\operatorname{Tr}\left(\Sigma^2\right) \Phi^{\dagger} \Phi$. % by rescaling the corresponding couplings. 
{Also, the potential cubic terms in the first line of Eq.~\eqref{eq:intreal2hdmc} break the $Z_2^{\Sigma}$ symmetry: $\Sigma\to-\Sigma$, 
and are negligible for the $2\to 2$ scalar scattering in the high energy limit. }
Therefore, these terms play no roles in deriving the perturbative unitarity bounds.
Furthermore, $\lambda_{10}$ can be a complex parameter and should vanish when the $Z_2$ symmetry involving the two Higgs doublets, softly-broken or not, is imposed.

\subsection{Scattering matrix}\label{sec:2hdmrealsc}

Based on the two-particle bases given in Tables~\ref{tab:dp22},~\ref{tab:dpreal23}, and \ref{tab:dpreal33} classified according to the conserved quantum numbers $(Y,I)$, % have been provided in Tabs.~\ref{tab:dp22},~\ref{tab:dpreal23}, and \ref{tab:dpreal33}.
we can expand the general potential in Eq.~\eqref{eq:real2HDM} with the scalar components defined in Eq.~\eqref{eq:hrcompont} and obtain the following scattering matrices of given $(Y,I)$:
%Using Eq.~\eqref{eq:hrcompont}, we expand the scalar potential~\eqref{eq:real2HDM} with the components of the Higgs scalar and determine the quartic couplings
%for the scattering peocess with a given $(Y,~I)$. The scattering matrices are summarized as follows:
\begin{equation}\label{eq:realS00}
    16 \pi S_{(0,0)}=\left(\begin{array}{ccccc}
    3 \lambda_1 & 2 \lambda_3+\lambda_4 & 3 \lambda_6 & 3 \lambda_6^* & \sqrt{3}\lambda_8 \\
    2 \lambda_3+\lambda_4 & 3 \lambda_2 & 3 \lambda_7 & 3 \lambda_7^* & \sqrt{3}\lambda_9 \\
    3 \lambda_6^* & 3 \lambda_7^* & \lambda_3+2 \lambda_4 & 3 \lambda_5^* & \sqrt{3}\lambda_{10}^* \\
    3 \lambda_6 & 3 \lambda_7 & 3 \lambda_5 & \lambda_3+2 \lambda_4 & \sqrt{3}\lambda_{10} \\
    \sqrt{3}\lambda_8 & \sqrt{3}\lambda_9 & \sqrt{3}\lambda_{10} & \sqrt{3}\lambda_{10}^* & 5\lambda_{\Sigma}
    \end{array}\right)
\end{equation}
\begin{equation}\label{eq:realS01}
    16 \pi S_{(0,1)}=\left(\begin{array}{cccc}
    \lambda_1 & \lambda_4 & \lambda_6 & \lambda_6^* \\
    \lambda_4 & \lambda_2 & \lambda_7 & \lambda_7^* \\
    \lambda_6^* & \lambda_7^* & \lambda_3 & \lambda_5^* \\
    \lambda_6 & \lambda_7 & \lambda_5 & \lambda_3
    \end{array}\right)
\end{equation}
\begin{equation}
    16 \pi S_{(0,2)}=2\lambda_{\Sigma}
\end{equation}
\begin{equation}
    16 \pi S_{(1,\frac{1}{2})}=
    16 \pi S_{(1,\frac{3}{2})}=\left(\begin{array}{cc}
    \lambda_8 & \lambda_{10}^* \\
    \lambda_{10} & \lambda_9 
    \end{array}\right)
\end{equation}
\begin{equation}
    16 \pi S_{(2,0)}=\lambda_3-\lambda_4
\end{equation}
\begin{equation}
    16 \pi S_{(2,1)}=\left(\begin{array}{ccc}
    \lambda_1 & \lambda_5^* & \sqrt{2} \lambda_6^* \\
    \lambda_5 & \lambda_2 & \sqrt{2} \lambda_7 \\
    \sqrt{2} \lambda_6 & \sqrt{2} \lambda_7^* & \lambda_3+\lambda_4
    \end{array}\right)
\end{equation}

Comparing with the 2HDM results in Ref.~\cite{Ginzburg:2005dt}, the scattering matrix $16\pi S_{(0,0)}$ now becomes 5-dimensional, since there is an additional state with $(Y,I)=(0,0)$ composed solely by components in the triplet $\Sigma$. %that consists of two components of the real triplet with $(Y=0,~I=0)$ (see Tab.~\ref{tab:dpreal33}) can scatter into the state
%that consists of two components of the Higgs doublet with the same hypercharge and isospin. 
Furthermore, the scattering processes in the sectors with $(Y,I)=(0,2),~(1,\frac{1}{2})$, and $(1,\frac{3}{2})$ take place only between two scalar triplets.

\subsection{$Z_2$ symmetry}

Now we simplify our discussion by imposing the softly broken $Z_2$ symmetry with $\Phi_1 \to \Phi_1$ and $\Phi_2 \to -\Phi_2$ on the 
scalar potential~\eqref{eq:real2HDM}, so that we have $\lambda_6=\lambda_7=\lambda_{10}=0$ but leaving a nonzero $m_{12}^2$.
%The softly broken $Z_2$ symmetry in the two Higgs doublets 
Such a model is phenomenologically important because it protects the 
theory from flavor changing neutral currents at tree level.
Using the results given in Appendix~\ref{apd:mevs}, the matrix~\eqref{eq:realS00} can be block diagonalized into a $2\times 2$ matrix $16\pi S_{(0,0)}^{(2)}$ and a $3\times 3$ one $16\pi S_{(0,0)}^{(3)}$  as follows
\begin{equation}\label{eq:real003}
    16\pi S_{(0,0)}^{(2)}=\left(\begin{array}{cc}
    \lambda_3+2 \lambda_4 & 3 \lambda_5^* \\
    3 \lambda_5 & \lambda_3+2 \lambda_4
    \end{array}\right)\,,\,\,
    16\pi S_{(0,0)}^{(3)}=\left(\begin{array}{ccc}
    3 \lambda_1 & 2 \lambda_3+\lambda_4 & \sqrt{3}\lambda_8 \\
    2 \lambda_3+\lambda_4 & 3 \lambda_2 & \sqrt{3}\lambda_9 \\
    \sqrt{3}\lambda_8 & \sqrt{3}\lambda_9 & 5\lambda_{\Sigma}
    \end{array}\right)\,.
\end{equation}
The eigenvalues for $16\pi S_{(0,0)}^{(3)}$ can be found numerically or analytically by applying Eq.~\eqref{eq:f3}.
Furthermore, the matrix $16 \pi S_{(0,1)}$ in Eq.~\eqref{eq:realS01} can also be decomposed into the following two matrices, 
\begin{eqnarray}
16\pi S_{(0,1)}^{\text{u}} = \left(\begin{array}{cc}
	\lambda_1 & \lambda_4 \\
	\lambda_4 & \lambda_1 
	\end{array}\right)\,, \quad  
 16\pi S_{(0,1)}^{\text{d}} =  \left(\begin{array}{cc}
 	\lambda_3 & \lambda_5^* \\
 	\lambda_5 & \lambda_3 
 \end{array}\right)\,.
\end{eqnarray}
Apart from $16\pi S_{(0,0)}^{(3)}$, the eigenvalues for the scattering matrices are summarized as follows:
\begin{equation}
    \begin{aligned}\label{eq:2HDMEV}
    &\Lambda_{(0,0)}^{(2)\pm}=\lambda_3+2 \lambda_4 \pm 3\left|\lambda_5\right|,\\
    &\Lambda_{(0,1)}^{\text{u}\pm}=\frac{1}{2}\left(\lambda_1+\lambda_2 \pm \sqrt{\left(\lambda_1-\lambda_2\right)^2+4 \lambda_4^2}\right),\\
    &\Lambda_{(0,1)}^{\text{d}\pm}=\lambda_3 \pm\left|\lambda_5\right|,\\
    &\Lambda_{(2,0)}=\lambda_3-\lambda_4,~~\Lambda_{(2,1)}=\lambda_3+\lambda_4,\\
    &\Lambda_{(2,1)}^{\pm}=\frac{1}{2}\left(\lambda_1+\lambda_2 \pm 
    \sqrt{\left(\lambda_1-\lambda_2\right)^2+4\left|\lambda_5\right|^2}\right),\\
    &\Lambda_{(0,2)}=2\lambda_{\Sigma},~~
    \Lambda_{(1,\frac{1}{2})}^1=\Lambda_{(1,\frac{3}{2})}^1=\lambda_8,~~
    \Lambda_{(1,\frac{1}{2})}^2=\Lambda_{(1,\frac{3}{2})}^2=\lambda_9,\\
    \end{aligned}
\end{equation}
where $\Lambda_{(0,0)}^{2\pm}$, $\Lambda_{(0,1)}^{\text{u}\pm}$, and $\Lambda_{(0,1)}^{\text{d}\pm}$ are the eigenvalues for 
$16\pi S_{(0,0)}^{(2)}$, $16\pi S_{(0,1)}^{\text{u}}$, and $16\pi S_{(0,1)}^{\text{d}}$.
By further assuming $\lambda_8=\lambda_9=\lambda_{\Sigma}=0$ in the matrix $S_{(0,0)}^{(3)}$, we can reproduce the 
eigenvalues $\Lambda_{00 \pm}^{\text {even }}$ in Eq.~(10) of Ref.~\cite{Ginzburg:2005dt}.
The last line of Eq.~\eqref{eq:2HDMEV} gives the eigenvalues for the scattering matrices involving only components of the real triplet.
Together with numerical eigenvalues of $16\pi S_{(0,0)}^{(3)}$, we have provided all eigenvalues for the model~\eqref{eq:real2HDM} with the softly broken $Z_2$ symmetry.

\subsection{The $\Sigma$SM model}

The $\Sigma $SM model is a simple extension of the SM by a real triplet scalar, many aspects of which has been extensively investigated in the literature, such as 
%The phenomenology of the extension of the SM with a real triplet scalar ($\Sigma$SM) have been investigated extensively in the previous literature, 
%including 
the dark matter phenomenology~\cite{FileviezPerez:2008bj,YaserAyazi:2014jby,Chiang:2020rcv}, the LHC searches~\cite{Wang:2013jba,Bandyopadhyay:2014vma}, 
and the strongly first-order EW phase transition~\cite{Niemi:2018asa,Bell:2020gug}. 
% Now we only have one Higgs doublet $\Phi$ which should be identified as the $\Phi_1$ in the 2HDM case. 
The potential of $\Sigma$SM can be reproduced by setting all the couplings in the 2HDM scalar potential
Eq.~\eqref{eq:real2HDM} involving the second Higgs doublet $\Phi_2$ to vanish, which is given by
%except $\lambda_1$, $\lambda_8$, and $\lambda_{\Sigma}$. In addition, we take $\lambda_1=2\lambda_{\Phi}$ to reproduce the commonly seen SM Higgs potential
\begin{equation}
    V_{\Sigma \rm SM} = V(\Phi)+V(\Sigma)+ V(\Phi, \Sigma)\,,
 \end{equation}
where $\Phi$ is the SM Higgs doublet, $V(\Sigma)$ is given in Eq.~\eqref{VSigma}, and
\begin{eqnarray}
    \label{eq:SMHP1}
    V(\Phi) &=& \mu^2\Phi^{\dagger} \Phi+\lambda_{\Phi}\left(\Phi^{\dagger} \Phi\right)^2\,, \\
    \label{eq:SMHP2}
    V(\Phi,\Sigma) &=& \frac{1}{\sqrt{2}} a_1 \Phi^{\dagger} \Sigma \Phi + \frac{\lambda_8}{2} \left( {\rm Tr} \Sigma^2 \right) \Phi^\dagger \Phi\,.
\end{eqnarray}
By using Eqs.~\eqref{eq:real003} and \eqref{eq:2HDMEV}, the unitarity bounds on the $\Sigma$SM are then found to be
\begin{equation}
    \begin{aligned}
    &|\lambda_{\Phi}|\leq 4\pi,~|\lambda_{\Sigma}|\leq 4\pi,~|\lambda_8|\leq 8\pi, \\
    &|6 \lambda_{\Phi}+5 \lambda_{\Sigma}\pm \sqrt{\left(6 \lambda_{\Phi}-5 \lambda_{\Sigma}\right)^2+12 \lambda_8}|\leq 16 \pi\,,
    \end{aligned}   
\end{equation}
which confirm the unitarity bounds provided in Ref.~\cite{Khan:2016sxm}.

\section{Two-Higgs-doublet model plus a complex triplet}\label{sec:2HDMcomplex}

In this section, we will consider the extension of 2HDM by a complex Higgs triplet $\Delta$ with $Y=2$~\cite{Chen:2021jok}. By using the state bases provided in Sec.~\ref{sec:Bases}, we shall calculate the scattering matrix for the most general case of the model. %Using these results, 
We shall then impose the perturbative unitarity constraints on the eigenvalues of the scattering matrix for several simplified models, such as
the one with a softly broken $Z_2$ symmetry and the Type-II seesaw model~\cite{Konetschny:1977bn,Cheng:1980qt,Magg:1980ut,Schechter:1980gr,Lazarides:1980nt,Mohapatra:1980yp,Mohapatra:1999zr}.

\subsection{The general scalar potential} 
The general scalar potential for the model with two Higgs doublets and a complex Higgs triplet scalar $\Delta$ %under the $SU(2)_L\times U(1)_Y$ gauge symmetry 
is given by
\begin{equation}\label{eq:2HDMCT}
    V_c=V\left(\Phi_1, \Phi_2\right)+V(\Delta)+V\left(\Phi_1, \Phi_2, \Delta\right),
\end{equation}
where the complex Higgs triplet is written as
\begin{equation}\label{eq:ctriplet}
    \Delta=\left(\begin{array}{cc}
    \delta^{+} / \sqrt{2} & \delta^{++} \\
    \delta^0 & -\delta^{+} / \sqrt{2}
    \end{array}\right)\,.
\end{equation}
Note that the neutral component $\delta^0$ is a complex scalar. The 2HDM potential $V\left(\Phi_1, \Phi_2\right)$ has been provided in Eq.~\eqref{eq:2HDM}, while
the part related to the self-interactions of the complex Higgs triplet is given by
\begin{equation}
    \label{eq:VDelta}
    V(\Delta)=m_{\Delta}^2 \operatorname{Tr} \Delta^{\dagger} \Delta+\lambda_{\Delta 1}
    \left(\operatorname{Tr} \Delta^{\dagger} \Delta\right)^2+\lambda_{\Delta 2} \operatorname{Tr}\left(\Delta^{\dagger} \Delta\right)^2.
\end{equation}
The third part in Eq.~\eqref{eq:2HDMCT} gives the interactions among the Higgs doublets and the triplet~\cite{Chen:2021jok}
\begin{equation}\label{eq:2hdmc3}
    \begin{aligned}
    V\left(\Phi_1, \Phi_2, \Delta\right)=&\left(\mu_1 \Phi_1^T i \tau_2 \Delta^{\dagger} \Phi_1
    +\mu_2 \Phi_2^T i \tau_2 \Delta^{\dagger} \Phi_2+\mu_3 \Phi_1^T i \tau_2 \Delta^{\dagger} \Phi_2+\text {H.c.}\right) \\
    &+\left[\lambda_8 \Phi_1^{\dagger} \Phi_1+\lambda_9 \Phi_2^{\dagger} \Phi_2+\left(\lambda_{10} \Phi_1^{\dagger} \Phi_2
    +\text {H.c.}\right)\right] \operatorname{Tr} \Delta^{\dagger} \Delta \\
    &+\lambda_{11} \Phi_1^{\dagger} \Delta \Delta^{\dagger} \Phi_1+\lambda_{12} \Phi_2^{\dagger} \Delta \Delta^{\dagger} \Phi_2
    +\left(\lambda_{13} \Phi_1^{\dagger} \Delta \Delta^{\dagger} \Phi_2+\text {H.c.}\right).
    \end{aligned}
\end{equation}
The $Z_2^{\Delta}$ symmetry of the transformation $\Delta\to -\Delta$ is only softly broken by the cubic terms in the first line of Eq.~\eqref{eq:2hdmc3}, which are negligible for the $2\to 2$ scalar scatterings in the high energy limit, so that they cannot be constrained by the unitarity bounds. {Note that there are possibly additional cubic interactions like $\Delta^a_b \Delta^b_c \Delta^c_a$. But we ignore them since they do not contribute to the unitarity bounds. }
The parameters $\lambda_{10}$ and $\lambda_{13}$ can be complex, and the associated interactions explicitly break the $Z_2$ symmetry involved in the two Higgs doublets. 

\subsection{Scattering matrix}\label{sec:2hdmcomplexsc}

We expand the scalar potential~\eqref{eq:2HDMCT} in terms of components in the two doublets and the triplet. With the two-particle eigenstates given in Tables.~\ref{tab:dp22},~\ref{tab:dpcomplex23}, and \ref{tab:dpcomplex33}, we can determine the scattering matrices for different conserved quantum numbers $(Y,I)$, which are summarized as follows:
\begin{equation}\label{eq:complexS00}
    16 \pi S_{(0,0)}=\left(\begin{array}{ccccc}
    3 \lambda_1 & 2 \lambda_3+\lambda_4 & 3 \lambda_6 & 3 \lambda_6^* & \lambda_a \\
    2 \lambda_3+\lambda_4 & 3 \lambda_2 & 3 \lambda_7 & 3 \lambda_7^* & \lambda_b \\
    3 \lambda_6^* & 3 \lambda_7^* & \lambda_3+2 \lambda_4 & 3 \lambda_5^* & \lambda_c^* \\
    3 \lambda_6 & 3 \lambda_7 & 3 \lambda_5 & \lambda_3+2 \lambda_4 & \lambda_c \\
    \lambda_a & \lambda_b & \lambda_c & \lambda_c^* & \lambda_{\Delta}
    \end{array}\right)~{\rm with}~
    \left\{\begin{matrix}
        \lambda_a&=&\sqrt{\frac{3}{2}}(2\lambda_8+\lambda_{11})\\ 
        \lambda_b&=&\sqrt{\frac{3}{2}}(2\lambda_9+\lambda_{12})\\ 
        \lambda_c&=&\sqrt{\frac{3}{2}}(2\lambda_{10}+\lambda_{13})\\
        \lambda_{\Delta}&=&2(4\lambda_{\Delta 1}+3\lambda_{\Delta 2})
    \end{matrix}\right.
\end{equation}
\begin{equation}\label{eq:complexS01}
    16 \pi S_{(0,1)}=\left(\begin{array}{ccccc}
    \lambda_1 & \lambda_4 & \lambda_6 & \lambda_6^* & \lambda_{11} \\
    \lambda_4 & \lambda_2 & \lambda_7 & \lambda_7^* & \lambda_{12} \\
    \lambda_6^* & \lambda_7^* & \lambda_3 & \lambda_5^* & \lambda_{13}^* \\
    \lambda_6 & \lambda_7 & \lambda_5 & \lambda_3 & \lambda_{13} \\
    \lambda_{11} & \lambda_{12} & \lambda_{13} & \lambda_{13}^* & 2\lambda_{\Delta 1}+4\lambda_{\Delta 2}
    \end{array}\right)
\end{equation}
\begin{equation}\label{eq:complexS02}
    16 \pi S_{(0,2)}=2\lambda_{\Delta 1}
\end{equation}
\begin{equation}
    16 \pi S_{(1,\frac{1}{2})}=\left(\begin{array}{cc}
    \lambda_8+3\lambda_{11}/2 & \lambda_{10}+3\lambda_{13}/2 \\
    \lambda_{10}^*+3\lambda_{13}^*/2 & \lambda_{9}+3\lambda_{12}/2
    \end{array}\right)
\end{equation}
\begin{equation}
    16 \pi S_{(1,\frac{3}{2})}=\left(\begin{array}{cc}
    \lambda_8 & \lambda_{10} \\
    \lambda_{10}^* & \lambda_9 
    \end{array}\right)
\end{equation}
\begin{equation}
    16 \pi S_{(2,0)}=\lambda_3-\lambda_4
\end{equation}
\begin{equation}
    16 \pi S_{2,1}=\left(\begin{array}{ccc}
    \lambda_1 & \lambda_5^* & \sqrt{2} \lambda_6 \\
    \lambda_5 & \lambda_2 & \sqrt{2} \lambda_7^* \\
    \sqrt{2} \lambda_6^* & \sqrt{2} \lambda_7 & \lambda_3+\lambda_4
    \end{array}\right)
\end{equation}
\begin{equation}
    16 \pi S_{(3,\frac{1}{2})}=\left(\begin{array}{cc}
    \lambda_8-\lambda_{11}/2 & \lambda_{10}^*-\lambda_{13}^*/2 \\
    \lambda_{10}-\lambda_{13}/2 & \lambda_{9}-\lambda_{12}/2
    \end{array}\right)
\end{equation}
\begin{equation}
    16 \pi S_{(3,\frac{3}{2})}=\left(\begin{array}{cc}
    \lambda_8+\lambda_{11} & \lambda_{10}^*+\lambda_{13}^* \\
    \lambda_{10}+\lambda_{13} & \lambda_{9}+\lambda_{12}
    \end{array}\right)
\end{equation}
\begin{equation}
    16 \pi S_{(4,0)}=2\lambda_{\Delta 1}-\lambda_{\Delta 2}
\end{equation}
\begin{equation}\label{eq:complexS42}
    16 \pi S_{(4,2)}=2(\lambda_{\Delta 1}+\lambda_{\Delta 2})
\end{equation}

{We observe that the scattering matrices $16\pi S_{(0,0)}$ and $16\pi S_{(0,1)}$ are now five-dimensional, which is compared with the four-dimensional matrices in the 2HDM.}
This is caused by the fact that irreducible representations of the product of two Higgs triplets contain the states with $(Y,I)=(0,0)$ and $(0,1)$, which can scatter into  
two components of Higgs doublets with the same quantum numbers. On the other hand,
the scattering processes with $(Y,I)=(0,2),~(1,\frac{1}{2}),~(1,\frac{3}{2})$, $(3,\frac{1}{2}),~(3,\frac{3}{2})$, $(4,0)$, and $(4,2)$ 
take place only among the complex Higgs triplets. 
% We note that since the positivity of $\lambda_{\Delta 1}$ and $\lambda_{\Delta 2}$ is not determined,
% the unitarity bounds from $16 \pi S_{(0,2)}$, $16 \pi S_{(4,0)}$, and $16 \pi S_{(4,2)}$ should be independently taken into account.

\subsection{Special case with a softly broken $Z_2$ symmetry}
Now we consider some simplified models in the above complex triplet extension of the 2HDM, which may allow us to obtain the eigenvalues of scattering matrices analytically. 
{The first example is to impose a softly broken $Z_2$ symmetry on the potential of the two Higgs doublets 
in Eq.~\eqref{eq:2HDMCT} so that we have the condition $\lambda_6=\lambda_7=\lambda_{10}=\lambda_{13}=0$ for the potential.}
In this case, the $5\times 5$ scattering matrix~\eqref{eq:complexS00} and~\eqref{eq:complexS01} can be decomposed into a 2-dimensional
(which is correspondingly denoted as $16\pi S_{(0,0)}^{(2)}$ and $16\pi S_{(0,1)}^{(2)}$) and a 3-dimensional matrices. 
The corresponding 2-dimensional and 3-dimensional matrices are given by
\begin{equation}\label{eq:complexM22}
    16\pi S_{(0,0)}^{(2)}=\left(\begin{array}{cc}
        \lambda_3+2 \lambda_4 & 3 \lambda_5^* \\
        3 \lambda_5 & \lambda_3+2 \lambda_4
        \end{array}\right),~
    16\pi S_{(0,1)}^{(2)}=\left(\begin{array}{cc}
      \lambda_3 & \lambda_5^* \\
      \lambda_5 & \lambda_3 
    \end{array}\right).
\end{equation}
\begin{equation}\label{eq:complexM33}
    16\pi S_{(0,0)}^{(3)}=\left(\begin{array}{ccc}
    3 \lambda_1 & 2 \lambda_3+\lambda_4 & \lambda_a \\
    2 \lambda_3+\lambda_4 & 3 \lambda_2 & \lambda_b \\
    \lambda_a & \lambda_b & \lambda_{\Delta}
    \end{array}\right),~
    16\pi S_{(0,1)}^{(3)}=\left(\begin{array}{ccc}
        \lambda_1 & \lambda_4 & \lambda_{11} \\
        \lambda_4 & \lambda_2 & \lambda_{12} \\
        \lambda_{11} & \lambda_{12} & 2\lambda_{\Delta 1}+4\lambda_{\Delta 2}
    \end{array}\right).
\end{equation}
It is convenient to find the eigenvalues for $16\pi S_{(0,0)}^{(3)}$ and $16\pi S_{(0,1)}^{(3)}$ numerically, 
we also provide the analytical solutions in Eq.~\eqref{eq:f3}.
We collect the eigenvalues for the remaining scattering matrices as follows:
% The eigenvalues for the two 2 dimension matrix $Y_{(0,0)}^{2}$ and $Y_{(0,1)}^{2}$ as well as the 
% matrix~\eqref{eq:complexS02}-\eqref{eq:complexS42} are summarized as follows:
\begin{equation}\label{eq:complexZ2}
    \begin{aligned}
    &\Lambda_{(0,0)}^{2\pm}=\lambda_3+2 \lambda_4 \pm 3\left|\lambda_5\right|,\\
    &\Lambda_{(0,1)}^{2\pm}=\lambda_3 \pm\left|\lambda_5\right|,\\
    &\Lambda_{(0,2)}=2\lambda_{\Delta 1},\\
    &\Lambda_{(1,\frac{1}{2})}^1=\lambda_8+3\lambda_{11}/2,~~\Lambda_{(1,\frac{1}{2})}^2=\lambda_9+3\lambda_{12}/2,\\
    &\Lambda_{(1,\frac{3}{2})}^1=\lambda_8,~~\Lambda_{(1,\frac{3}{2})}^2=\lambda_9,\\
    &\Lambda_{(2,0)}=\lambda_3-\lambda_4,~~\Lambda_{(2,1)}^{1}=\lambda_3+\lambda_4,\\
    &\Lambda_{(2,1)}^{2\pm}=\frac{1}{2}\left(\lambda_1+\lambda_2 \pm 
    \sqrt{\left(\lambda_1-\lambda_2\right)^2+4\left|\lambda_5\right|^2}\right),\\
    &\Lambda_{(3,\frac{1}{2})}^1=\lambda_8-\lambda_{11}/2,~~\Lambda_{(3,\frac{1}{2})}^2=\lambda_9-\lambda_{12}/2,\\
    &\Lambda_{(3,\frac{3}{2})}^1=\lambda_8+\lambda_{11},~~\Lambda_{(3,\frac{3}{2})}^2=\lambda_9+\lambda_{12},\\
    &\Lambda_{(4,0)}=2\lambda_{\Delta 1}-\lambda_{\Delta 2},~~\Lambda_{(4,2)}=2(\lambda_{\Delta 1}+\lambda_{\Delta 2}),
    \end{aligned}
\end{equation}
where $\Lambda_{(0,0)}^{2\pm}$ and $\Lambda_{(0,1)}^{2\pm}$ are the eigenvalues for $16\pi S_{(0,0)}^{(2)}$ and $16\pi S_{(0,1)}^{(2)}$, respectively.
The remaining results in Eq.~\eqref{eq:complexZ2} represent the eigenvalues for the matrices~\eqref{eq:complexS02}-\eqref{eq:complexS42}.
Combining with numerical eigenvalues of the matrices~$16\pi S_{(0,0)}^{(3)}$ and $16\pi S_{(0,1)}^{(3)}$, we have provided all eigenvalues for the scattering
matrices in the model~\eqref{eq:2HDMCT} with a softly broken $Z_2$ symmetry. 
Note that the unitarity bounds of the same model was already considered in Ref.~\cite{Ouazghour:2018mld}. 
Although we have used different notations to parametrize the scalar potential from Ref.~\cite{Ouazghour:2018mld}, 
a careful comparison of their eigenvalues of the scattering matrices given in Eqs.~(40) and (41) of Ref.~\cite{Ouazghour:2018mld} with the 
counterparts in Eqs.~\eqref{eq:complexM33} and \eqref{eq:complexZ2} shows that most of them are actually exactly the same. 
The only exception is that authors of Ref.~\cite{Ouazghour:2018mld} seemed to miss the cubic equation that corresponds to the 
first matrix in Eq.~\eqref{eq:complexM33} of our work.

\subsection{The $\Delta $SM}
Another simple example belonging the present class is to consider a model with only one Higgs doublet and one complex triplet, the so-called $\Delta$SM, which
%The extension of SM with a complex triplet ($\Delta $SM) 
has been widely employed to explain the tiny neutrino masses by the type-II 
seesaw mechanism~\cite{Konetschny:1977bn,Cheng:1980qt,Magg:1980ut,Schechter:1980gr,Lazarides:1980nt,Mohapatra:1980yp,Mohapatra:1999zr,Gu:2006wj,Chao:2007mz,FileviezPerez:2008jbu}. The investigations of this model are extended to the 
searches at colliders~\cite{Melfo:2011nx,Chen:2013dh,Han:2015sca,Dev:2018sel,Du:2018eaw,Cheng:2022jyi}, dark matter~\cite{Ding:2017jdr} and electroweak phase transition (EWPT) phenomena~\cite{Arhrib:2011uy,Primulando:2019evb,Zhou:2022mlz}.
%From potential~\eqref{eq:2HDMCT}, we can reproduce the Higgs sector of $\Delta $SM by keeping the quartic terms with couplings 
%$\lambda_1$, $\lambda_8$, $\lambda_{11}$, $\lambda_{\Delta 1}$, and $\lambda_{\Delta 2}$ while setting all other quartic couplings to zero.
{The scalar potential for $\Delta $SM is given by
\begin{equation}
    V_{\Delta \rm SM} = V(\Phi)+V(\Delta)+ V(\Phi, \Delta)\,,
\end{equation}
where $V(\Phi)$ is the SM Higgs potential in Eq.~\eqref{eq:SMHP1}, $V(\Delta)$ is given by Eq.~\eqref{eq:VDelta}, and
\begin{equation}
    V(\Phi, \Delta)=\left( \mu_1 \Phi^T i \tau_2 \Delta^{\dagger} \Phi + {\rm H.c.}\right) 
    + \lambda_8 \Phi^{\dagger} \Phi\operatorname{Tr} \Delta^{\dagger} \Delta
    +\lambda_{11} \Phi^{\dagger} \Delta \Delta^{\dagger} \Phi.
\end{equation}
}

% {\color{red} We also replace $\lambda_1$ in potential~\eqref{eq:2HDMCT} by $2\lambda_{\Phi}$ to reproduce the commonly seen SM Higgs potential given by Eq.~\eqref{eq:SMHP1}.} 
By using the matrices in Eq.~\eqref{eq:complexM33} and the eigenvalues in Eq.~\eqref{eq:complexZ2}, the unitarity bounds for the $\Delta $SM are given by
\begin{equation}
    \begin{aligned}
    &|\lambda_{\Phi}|\leq 4\pi,~|\lambda_{\Delta 1}|\leq 4\pi,~|\lambda_8|\leq 8\pi\\
    &|2\lambda_{\Delta 1}-\lambda_{\Delta 2}|\leq 8\pi,~|\lambda_{\Delta 1}+\lambda_{\Delta 2}|\leq 4\pi,\\
    &|\lambda_8+3\lambda_{11}/2|\leq 8\pi,~|\lambda_8-\lambda_{11}/2|\leq 8\pi,~|\lambda_8+\lambda_{11}|\leq 8\pi,\\
    &|\lambda_{\Phi}+\lambda_{\Delta 1}+2\lambda_{\Delta 2}\pm \sqrt{(\lambda_{\Phi}-\lambda_{\Delta 1}-2\lambda_{\Delta 2})+\lambda_{11}^2}|\leq 8\pi,\\
    &|6\lambda_{\Phi}+8\lambda_{\Delta 1}+6\lambda_{\Delta 2}\pm \sqrt{(6\lambda_{\Phi}-8\lambda_{\Delta 1}-6\lambda_{\Delta 2})^2+
    6(2\lambda_8+\lambda_{11})^2}|\leq 16\pi\,,
    \end{aligned}
\end{equation}
which are in agreement with the results given in Ref.~\cite{Arhrib:2011uy}.

\section{Applications}\label{sec:application}

The unitarity constraint on the quartic couplings can be translated into the upper bounds on the Higgs boson masses if $\sqrt{\lambda_i}v$
dominates the masses of the associated Higgs bosons. 
{The unitarity bound as well as other constraints on the 2HDM have been fully explored in previous literature (see {\it e.g.} Ref.~\cite{Branco:2011iw} for a review and references there in). Moreover, we would like to mention that it is valuable to investigate the phenomenology of the real triplet extension of the 2HDM, which has been somewhat less explored in the literature. However, the careful study requires not only the unitarity bounds derived in Sec.~\ref{sec:2HDMreal} but also many other theoretical and experimental constraints, which are obviously beyond the scope of the present work. Thus, in this section, we shall focus on the perturbative unitarity bounds to the complex triplet extension of the 2HDM, and show the quantitative constraints on the model parameters. 
Note that this model was recently proposed to explain the muon $g-2$ anomaly in Ref.~\cite{Chen:2021jok}. }

The muon anomalous magnetic dipole moment (denoted by $(g-2)_{\mu}$) is one of the long-standing anomalies in the particle physics. {This discrepancy has further been confirmed by the recent muon $g-2$ measurement performed by the Muon experiment at Fermilab, which has yielded the most precise experimental muon $g-2$ value $a_\mu^{\rm Exp} = (116\,592\,061\pm 41)\times 10^{-11}$~\cite{Muong-2:2021ojo} by combining the Brookhaven data~\cite{Muong-2:2006rrc} . On the other hand, the state-of-the-art calculations of various SM contributions~\cite{Aoyama:2012wk,Aoyama:2019ryr,Czarnecki:2002nt,Gnendiger:2013pva,Davier:2017zfy,Keshavarzi:2018mgv,Colangelo:2018mtw,Hoferichter:2019mqg,Davier:2019can,Keshavarzi:2019abf,Kurz:2014wya,Melnikov:2003xd,Masjuan:2017tvw,Colangelo:2017fiz,Hoferichter:2018kwz,Gerardin:2019vio,Bijnens:2019ghy,Colangelo:2019uex,Blum:2019ugy,Colangelo:2014qya} predict $a_\mu^{\rm SM} = (116\,591\,810\pm 43)\times 10^{-11}$ (see {\it e.g.} Ref.~\cite{Aoyama:2020ynm} for a recent review and reference therein). As a result, the discrepancy between the SM and experimental values of $a_\mu$ is given by~\cite{Muong-2:2021ojo}
\begin{eqnarray}
	\Delta a_\mu = a_\mu^{\rm Exp} - a_\mu^{\rm SM} = (251\pm 59)\times 10^{-11}\,,
\end{eqnarray}
with the significance reaching $4.25\sigma$.} Possible solutions to the muon $g-2$ anomaly has been widely discussed in the 2HDM content in Refs.~\cite{Cheung:2001hz,Cheung:2003pw,Zhou:2001ew,Aoki:2009ha,Cao:2009as,Han:2022juu}.
At one-loop level, both the charged and neutral Higgs bosons in the 2HDM contribute to the muon $g-2$, 
but it is found that these corrections are too small to explain the observed deviation.
On the other hand, the two-loop Barr-Zee diagrams %in which the scalar or pseudo-scalar couples to a heavy fermion loop 
can give rise to the dominant
contribution to the muon $g-2$ in some parameter space. However, it has been shown that the explanation of the muon $g-2$ anomaly with the Barr-Zee mechanism requires a light pseudo-scalar
mass with $m_{A}\lesssim 100$~GeV and $t_{\beta}\sim 50$ when various constraints are imposed~\cite{Cheung:2003pw,Ferreira:2021gke,Kim:2022xuo}.
Note that one class of the strictest constraints is provided by the unitarity bounds in the theory. In particular, Ref.~\cite{Ferreira:2021gke} has shown 
that most of the parameter space with $m_{A}\gtrsim 100$~GeV in the typical 2HDM is already excluded by the unitarity alone. 

%*****************************figure3***************************************
\begin{figure}
    \centering
    \includegraphics[width=70mm,angle=0]{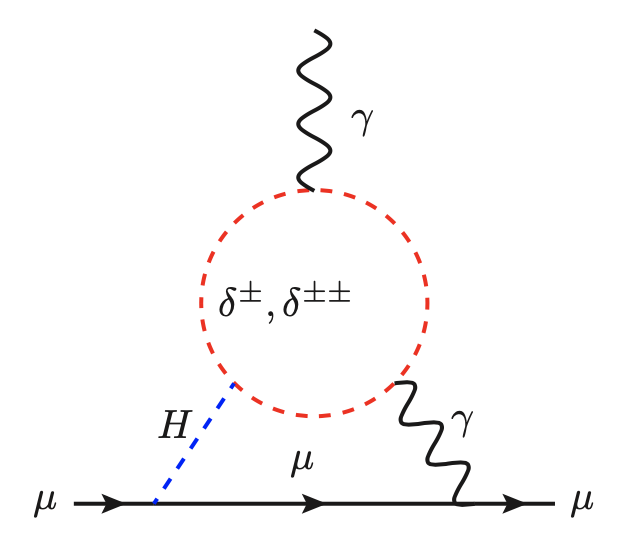} 
    \caption{The Barr-Zee type Feynman diagrams for the muon $g-2$, with charged scalars $\delta^{\pm}$ and $\delta^{\pm\pm}$ running in the loops.} 
    \label{fig:BZ}
  \end{figure} 
%*************************************************************************

%*****************************figure4***************************************
\begin{figure}
    \centering
    \includegraphics[width=75mm,angle=0]{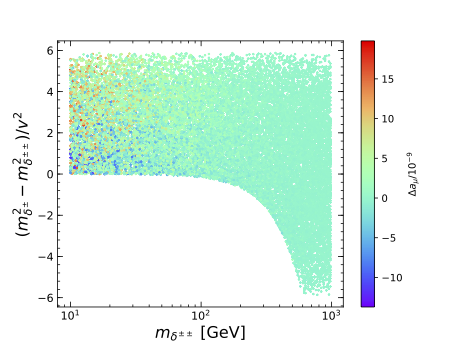}
    \includegraphics[width=75mm,angle=0]{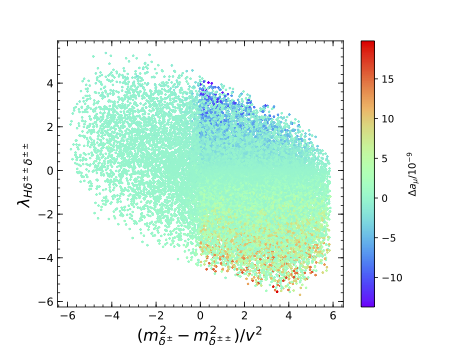}\\
    \includegraphics[width=75mm,angle=0]{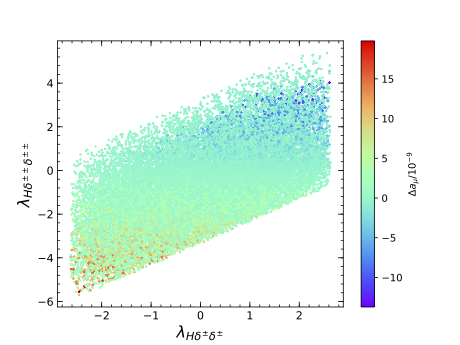}
    \includegraphics[width=75mm,angle=0]{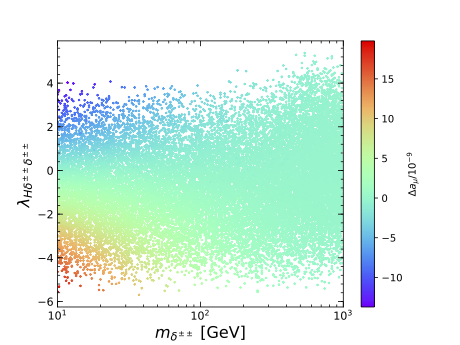}
    \caption{Unitarity bounds on the relevant parameter spaces of the complex triplet extension of the 2HDM. % the mass difference of $m_{\delta^{\pm}}-m_{\delta^{\pm\pm}}$. Lower: unitarity bound on the trilinear couplings
   % $\lambda_{H\delta^{\pm}\delta^{\pm}}$ and $\lambda_{H\delta^{\pm\pm}\delta^{\pm\pm}}$. 
    The colorbar represents the values of $\Delta a_{\mu}/10^{-9}$.}
    \label{fig:g21}
  \end{figure}
%*************************************************************************

The recent work in Ref.~\cite{Chen:2021jok} shows that if a complex Higgs triplet is added to the 2HDM, the charged components of the 
Higgs triplet can induce new Barr-Zee-type contributions illustrated in Fig.~\ref{fig:BZ}, which may explain the muon $g-2$ while easily 
evading other experimental constraints. % in which the fermion couples to the charged scalars running in the loop via the 
%neutral scalars. 
From these Feynman diagrams, it is clear that the new contribution to muon $g-2$ is proportional to the trilinear scalar couplings $\lambda_i v$ which might be well constrained by the 
perturbative unitarity. %can play an important role in constraining such a contribution. 
However, Ref.~\cite{Chen:2021jok} only applied the approximate unitarity bounds from the aligned two-Higgs doublet 
model (A2HDM)~\cite{Pich:2009sp}  to constrain the parameter space. It is more appropriate to apply the exact unitarity bounds derived 
in the previous section for this complex triplet extension of the 2HDM, which is the main motivation for this section.
	
	% Ref.~\cite{Chen:2021jok} has not appropriately taken into account the unitarity issue, which is the main topic in this subsection.}

Following Ref.~\cite{Chen:2021jok}, we will consider the decoupling limit of the model. By using the minimization conditions for the 
scalar potential in Eq.~\eqref{eq:2HDMCT} %{\color{blue} with respect to $d^0$ (where $\delta^0=\frac{1}{\sqrt{2}}(d^0+i\eta^0)$)} 
and the mass matrices provided in 
Appendix~\ref{apd:Mmatrix}, we find that the decoupling between components in Higgs doublets and those in the triplet can be achieved when $v_{\Delta},~\tilde{\mu}_3\ll 1$~GeV.
In the two-Higgs-doublet sector, we consider the case with a softly broken $Z_2$ symmetry and $CP$ conservation, in which all parameters in the scalar potential are real. 
Also, following Ref.~\cite{Chen:2021jok}, we shall consider the aligned limit of the two Higgs doublets, i.e., $c_{\beta-\alpha} \approx 0$.
In this case, the mass eigenstates $h$ and $H$ are almost $h_1^0$ and $h_2^0$, so that the trilinear scalar couplings are given by 
Eq.~\eqref{eq:tricp} in Appendix~\ref{apd:tricoupling}.

To calculate the Barr-Zee diagram shown in Fig.~\ref{fig:BZ}, we need to know the coupling between muons and $H$. 
As in Ref.~\cite{Chen:2021jok}, we shall consider the A2HDM case~\cite{Pich:2009sp} 
(see Ref.~\cite{Eberhardt:2020dat} for the recent global fit of A2HDM), in which 
the lepton Yukawa couplings with $H$ are given by
\begin{equation}
    -{\cal L}_{Y}=\sum_f y_{f}^H\frac{M_f}{v}\bar{f}_Lf_RH+{\rm H.c.},
\end{equation}
where $M_f$ is the mass of the lepton flavor $f$ and %the Yukawa coupling is given by 
\begin{equation}
    y_f^H = \left(s_{\beta-\alpha} \zeta_f-c_{\beta-\alpha}\right)\,.
\end{equation}
Here $\zeta_f$ is a parameter in the A2HDM, whose benchmark value is taken to be $\zeta_f=-100$ following Ref.~\cite{Chen:2021jok}.

The contribution to the muon $(g-2)$ from the Barr-Zee diagrams is given by~\cite{Ilisie:2015tra}
\begin{equation}\label{eq:dag2}
    \Delta a_\mu=\sum_{\phi_i} \frac{\alpha m_\mu^2}{8 \pi^3 m_{H}^2} \operatorname{Re}\left(y_f^{H}\right) 
    \lambda_{H \phi_i\phi_i^*} \mathcal{F}\left(\frac{m_{\phi_i}^2}{m_{H}^2}\right),
\end{equation}
where $\phi_i=\delta^{\pm},~\delta^{\pm\pm}$, the trilinear couplings $\lambda_{H \phi_i\phi_i^*}$ are given in Eq.~\eqref{eq:tricp}, 
and the loop function is given by 
\begin{equation}
    \mathcal{F}(\omega)=\frac{1}{2} \int_0^1 d x \frac{x(x-1)}{\omega-x(1-x)} \ln \left(\frac{\omega}{x(1-x)}\right).
\end{equation}
%To focus on our topic, 
Since the Barr-Zee diagrams in Fig.~\ref{fig:BZ} dominate the anomalous muon $g-2$, we can ignore other one- or two-loop $(g-2)_\mu$ contributions in our following numerical calculations. % only take into account the contributions from the Barr-Zee diagram given by  and neglect the other Feynman diagrams.

In order to search for the parameter space allowed by the perturbative unitarity, we scan over the quartic couplings $\lambda_8$, $\lambda_9$,
$\lambda_{11}$, and $\lambda_{12}$ in the range of $(-8\pi-8\pi)$ and the doubly-charged scalar mass $m_{\delta^{\pm\pm}}$ in the range of
$(10-1000)$~GeV. For the 2HDM sector, we take $\lambda_{1,2,...,5}=0.2$, $t_{\beta}=5$, and $m_{H}=300$~GeV for conservative estimations.
The other parameters in the model are set to zero.
{The relations among various parameters in the generic scalar basis and the Higgs basis are summarized in Appendix~\ref{apd:HBcouplings}, while
the masses of scalars are determined in Appendix~\ref{apd:Mmatrix}. }

We show the scan results in Fig.~\ref{fig:g21}. From the upper two plots, we observe that the mass squared difference between the singly-charged and doubly-charged scalars in the triplet should be 
$|m^2_{\delta^{\pm}}-m^2_{\delta^{\pm\pm}}|/v^2 \lesssim 6$, which is  restricted by unitarity bound on the quartic coupling $\tilde{\lambda}_{11}$ as seen in Eqs.~\eqref{eq:SCmass} and \eqref{eq:DCmass}.
%By using Eqs.~\eqref{eq:SCmass} and \eqref{eq:DCmass}, we see that the mass difference between the singly-charged and 
%doubly-charged scalars is determined by the quartic couplings $\tilde{\lambda}_{11}$, and therefore, is restricted by the unitarity.
Furthermore, for $m_{\delta^{\pm\pm}}\lesssim 200$~GeV we have $m^2_{\delta^{\pm}}-m^2_{\delta^{\pm\pm}}>0$. % which means that there is 
%a lower limit on the mass difference from the unitarity bound. 
The colorbar of this figure represents the distribution of the predicted $\Delta a_{\mu}$ values. The lower two plots of Fig.~\ref{fig:g21} show that large values of $|\Delta a_{\mu}|$ prefer large values of trilinear scalar couplings, $|\lambda_{H\delta^{\pm}\delta^{\pm}}|$ and 
$|\lambda_{H\delta^{\pm\pm}\delta^{\pm\pm}}|$, as well as small values of $m_{\delta^{\pm\pm}}$. Note that if $\Delta a_{\mu}$ is positive as required by experiments, it picks the parameter space with negative values of $\lambda_{H\delta^{\pm}\delta^{\pm}}$ and/or $\lambda_{H\delta^{\pm\pm}\delta^{\pm\pm}}$, which are well constrained by the unitarity consideration with 
%, the two trilinear scalar couplings are restricted within the ranges
 $|\lambda_{H\delta^{\pm}\delta^{\pm}}|\lesssim 2.6$ and 
$|\lambda_{H\delta^{\pm\pm}\delta^{\pm\pm}}|\lesssim 5.0$.  From the Barr-Zee Feynman diagrams and their expressions in Eq.~\eqref{eq:dag2}, %we see that for Barr-Zee diagrams, 
the trilinear couplings are directly related to the dominant contribution to the muon $g-2$, and perturbative unitarity can thus put very useful constraints on this model.
Finally, we note that the unitarity bounds given in this section are rather conservative since the doublet-triplet mixings are ignored due to the nearly vanishing triplet VEV. In the case with $v_\Delta \sim 1$~GeV, the mixings between Higgs doublet and triplet components can become significant, which would further enhance the unitarity bounds on the scalar masses. %As we have shown in the case of C2HDM, the constraints
%on the scalar masses from perturbative unitarity can be very strong. 

We make several final remarks before closing this subsection. The difference between Ref.~\cite{Chen:2021jok} and ours is obvious. 
Ref.~\cite{Chen:2021jok} aims to explain the muon $g-2$ in the context of the extension 2HDM with a complex triplet scalar. Our work focus
on the derivations of the unitarity bounds for this specific extension of 2HDM. In this section, we have applied the unitarity bounds 
to constrain the trilinear scalar couplings $\lambda_{H \delta^{\pm} \delta^{\pm}}$ and $\lambda_{H \delta^{\pm\pm}\delta^{\pm\pm}}$, 
which has not been done in Ref.~\cite{Chen:2021jok}. 
Our new findings are depicted in Fig.~\ref{fig:g21}, which indicates that the following parameter regions% the trilinear scalar couplings and the mass squared difference between charged triplet scalars  in the following ranges
\begin{equation}\label{eq:ubtls}
    |\lambda_{H\delta^{\pm}\delta^{\pm}}|\gtrsim 2.6,~~|\lambda_{H\delta^{\pm\pm}\delta^{\pm\pm}}|\gtrsim 5.0,~~{\rm and}~~
    |m_{\delta^{\pm}}^2-m_{\delta^{\pm\pm}}^2|\gtrsim 6v^2
\end{equation}
have been excluded by perturbative unitarity.
We also note that Ref.~\cite{Chen:2021jok} applied the value $\lambda_{H \delta^{\pm\pm}\delta^{\pm\pm}}=5$ for their estimations of muon $g-2$. 
This value just lies at the $\lambda_{H \delta^{\pm\pm}\delta^{\pm\pm}} $ upper limit allowed by the unitarity bounds in Eq.~\eqref{eq:ubtls}. 
Therefore, we conclude that the main conclusion drawn by Ref.~\cite{Chen:2021jok} that the 2HDM with a complex triplet can explain the muon $g-2$
does not change dramatically even if the unitarity bounds are appropriately addressed. 

\section{Conclusions}\label{sec:conclusions}

The perturbative unitarity is one of the most significant theoretical constraints on the Higgs sector, beyond which the perturbation calculation in the theory breaks down.
It has proven to be successful in predicting the upper limit on the Higgs boson mass in the minimal SM, and applying to constrain many new physics models such the 2HDM.
In this work, we focus on deriving the perturbative unitarity bounds on the extension of the 2HDM with an additional real or complex Higgs scalar triplet.
Since the total hypercharge and isospin are conserved in the high-energy limit of scatterings, we explicitly give the two-particle state basis according to 
their $SU(2)_L\times U(1)_Y$ charges by decomposing the direct product of two Higgs multiplets into direct sums of irreducible representations 
under electroweak gauge groups. The classification of the two-particle state basis is summarized in Tables~\ref{tab:dp22}-\ref{tab:dprealcomplex33}, 
in which the states are expressed in terms of component fields.
% decompose the direct product of two Higgs multiplets into its direct sum and classify
%the bases (which are summarized in Tabs.~\ref{tab:dp22}-\ref{tab:dprealcomplex33}) of the irreducible representations according the conserved quantum numbers. 
With these two-particle bases, the $2\to 2$ scattering amplitudes among scalars can be simplified into the block-diagonal forms, 
which are easily determined by expanding the quartic scalar terms in the potential. We then impose the unitarity bound on the eigenvalues 
of the scattering matrices. The associated analytical results are summarized in Secs.~\ref{sec:2hdmrealsc} and \ref{sec:2hdmcomplexsc}.

{We then numerically apply our derived unitarity bounds to the extension of 2HDM with a complex Higgs triplet, which was recently shown to be of great phenomenological interest.
We have shown that the unitarity can put strong upper limits on the trilinear scalar couplings and the mass differences of the charged triplet scalars.
Since the contributions to the  muon $g-2$ from the Barr-Zee diagrams with a charged scalar running in the loop are proportional to the trilinear 
scalar couplings, the unitarity bounds on these couplings can also constrain new solutions to the long-standing muon $g-2$ anomaly.}
In the near future, together with the experimental measurements of the Higgs trilinear coupling and the Higgs signal strengths of different 
channels at the LHC Run 3, we hope that the unitarity bounds would help us to understand the structure of Higgs sector more deeply.  

\section*{Acknowledgments}
BQL is supported in part by Zhejiang Provincial Natural Science Foundation of China under Grant No. LQ23A050002
and National Natural Science Foundation of China (NSFC) under Grant No. 12147219.
DH is supported in part by the National Natural Science Foundation of China (NSFC) under Grant No. 12005254, the National Key Research and Development Program of China under Grant No.
2021YFC2203003, and the Key Research Program of
Chinese Academy of Sciences under grant No. XDPB15

\appendix

\section{Eigenvalues of the $(Y,I)=(0,0)$ scattering matrix in the triplet extension of the 2HDM with a $Z_2$ symmetry}
\label{apd:mevs}

In this section we analytically solve the eigenvalues for the 5-dimensional scattering matrix, which appears in the $(Y,I)=(0,0)$ sector of the extension of 2HDM with a softly-broken $Z_2$ symmetry.
By imposing the $Z_2$ symmetry, the 5-dimensional scattering matrix in Eq.~\eqref{eq:complexS00} can be simplified into the following form
\begin{equation}
    X=\left(\begin{array}{ccccc}
    a_1 & a_2 & 0 & 0 & c_5 \\
    a_3 & a_4 & 0 & 0 & c_6 \\
    0 & 0  &  b_1 & b_2 & 0 \\
    0 & 0  &  b_3 & b_4 & 0 \\
    c_5 & c_6 & 0 & 0 & c_7
    \end{array}\right)\,,
\end{equation}
which can be further decomposed into a $2\times 2$ matrix and a $3\times 3$ one as follows
\begin{equation}
    X^{(2)}=\left(\begin{array}{cc}
        b_1 & b_2 \\
        b_3 & b_4 \\
    \end{array}\right),~~
    X^{(3)}=\left(\begin{array}{ccc}
    a_1 & a_2 & c_5 \\
    a_3 & a_4 & c_6 \\
    c_5 & c_6 & c_7
    \end{array}\right).
\end{equation}
The eigenvalues for $X^{(2)}$ and $X^{(3)}$ are the same as that directly obtained from the 5-dimensional matrix $X$.
Note that, for a general matrix $A$, the eigenvalue $f$ can be obtained by solving the equation
\begin{equation}
    |fI-A|=0\,.
\end{equation}
%where $f$ denotes the eigenvalues of the matrix $A$. 
For $X^{(2)}$ and $X^{(3)}$, the eigenvalue equation can be transformed into the following equations
\begin{equation}\label{eq:quadratic}
    (f-b_1)(f-b_4)-b_2b_3=0
\end{equation}
\begin{equation}\label{eq:cubic1}
    (f-a_1)(f-a_4)(f-c_7)-(f-a_1)c_6^2-a_2a_3(f-c_7)-a_2c_5c_6-a_3c_5c_6-(f-a_4)c_5^2=0\,,
\end{equation}
respectively. 
The solutions for Eq.~\eqref{eq:quadratic} can be easily solved by
\begin{equation}
    f_{1,2}=\frac{1}{2}\left( b_1+b_4\pm\sqrt{(b_1-b_4)^2+4b_2b_3} \right).
\end{equation}
For the case with $b_1=b_4$ and $b_2=b_3^*$, we have
\begin{equation}
    f_{1,2}= b_1\pm |b_2|.
\end{equation}
Note that the scattering matrix should be Hermitian, which means $a_2=a_3^*$ and $b_2=b_3^*$.
Furthermore, the eigenvalues for the Hermitian matrix are always real. Note that, for a general cubic equation $f^3+bf^2+cf+d=0$, one representations of the three roots is given by  
%the three roots for the cubic equation $f^3+bf^2+cf+d=0$ are given by
\begin{equation}\label{eq:f3}
    \begin{aligned}
    &f_3=-\frac{b}{3} + 2 \sqrt[3]{r} \cos \theta \,, \\
    &f_4=-\frac{b}{3} + 2 \sqrt[3]{r} \cos \left(\theta+\frac{2}{3}\pi\right)\,, \\
    &f_5=-\frac{b}{3} + 2 \sqrt[3]{r} \cos \left(\theta+\frac{4}{3}\pi\right)\,,
    \end{aligned}
\end{equation}
where
\begin{equation}
    r=\sqrt{-\left(\frac{p}{3}\right)^3},~\theta=\frac{1}{3} \arccos \left(-\frac{q}{2 r}\right),~{\rm with}~
    p=\frac{3c-b^2}{3},~q=\frac{27d-9bc+2b^3}{27}\,.
\end{equation}
By comparing Eq.~\eqref{eq:cubic1} with the general cubic equation, it is found that
\begin{equation}
    \begin{aligned}
    &b=-(a_1+a_4+c_7)\,,\\
    &c=a_1a_4+a_1c_7+a_4c_7-a_2a_3-c_5^2-c_6^2\,,\\
    &d=a_1c_6^2+a_2a_3c_7+a_4c_5^2-a_2c_5c_6-a_3c_5c_6\,.
    \end{aligned}
\end{equation}
In this way, we give the analytic solutions to the eigenvalues for the three-rank scattering matrix $X^{(3)}$.

\section{Parameters in the Higgs basis for the complex triplet extension of 2HDM}
\label{apd:HBcouplings} 

The electroweak gauge symmetry is spontaneously broken when the neutral components of the Higgs multiplets obtain VEVs.
The VEVs can be complex and there may be a relative phase between them. Here we use $\xi$ to denote the phase between the VEVs of 
doublets $\Phi_1$ and $\Phi_2$ in the triplet extension of the 2HDM. Concretely, one assumes real $v_1$ and complex $v_2e^{i\xi}$. 
Such a phase can be absorbed by the following phase redefinitions of the complex parameters:
\begin{equation}\label{eq:replace1}
    \lambda_5 \rightarrow e^{2 i \xi} \lambda_5\text{ and } 
    m_{12}^2, \lambda_6, \lambda_7, \lambda_{10}, \lambda_{13} \rightarrow 
    e^{i \xi}\left\{m_{12}^2, \lambda_6, \lambda_7, \lambda_{10}, \lambda_{13}\right\}.
\end{equation}
so that the form of the potential keeps unchanged. Thus, we can start with real VEVs for scalars.
We can rotate the generic scalar basis $\{\Phi_1,\Phi_2 \}$ into the Higgs basis $\{H_1,H_2 \}$ via the transformation
\begin{equation}
    \label{eq:trans1}
    \left(\begin{array}{c}
    H_{1} \\
    H_{2}
    \end{array}\right)=\left(\begin{array}{cc}
    \cos\beta & \sin\beta \\
    -\sin\beta & \cos\beta
    \end{array}\right)\left(\begin{array}{c}
    \Phi_{1} \\
    \Phi_{2}
    \end{array}\right)
\end{equation}
so that only $H_1$ has a non-vanishing VEV $v=\sqrt{v_1^2+v_2^2}\simeq 246$~GeV.
Here the quantity $\tan\beta$ is defined by the ratio of two Higgs field VEVs, i.e., $\tan\beta\equiv t_{\beta} =v_2/v_1$.
In the following, we summarize the parameters of the potential in the Higgs basis as functions of those defined in the generic basis.
For the parameters with mass dimensions, we have
\begin{eqnarray}\label{eq:dimHG}
    \tilde{m}_{11}^2&=&c_{\beta}^2m_{11}^2+s_{\beta}^2m_{2}^2-c_{\beta}s_{\beta}\left( m_{12}^2+m_{12}^{2*} \right),\nonumber\\
    \tilde{m}_{22}^2&=&s_{\beta}^2m_{11}^2+c_{\beta}^2m_{2}^2+c_{\beta}s_{\beta}\left( m_{12}^2+m_{12}^{2*} \right),\nonumber\\
    \tilde{m}_{12}^2&=&c_{\beta}s_{\beta}\left( m_{11}^2-m_{22}^{2} \right)+\left( c_{\beta}^2m_{12}^2-s_{\beta}^2m_{12}^{2*} \right),\nonumber\\
    \tilde{m}_{\Delta}^2&=&m_{\Delta}^2,\nonumber\\
    \tilde{\mu}_{1}&=&\mu _{1}c_{\beta }^2+\mu _{3}c_{\beta } s_{\beta }+\mu _{2} s_{\beta }^2,\nonumber\\
    \tilde{\mu}_{2}&=&\mu _{2}c_{\beta }^2-\mu _{3}c_{\beta } s_{\beta }+\mu _{1} s_{\beta }^2,\nonumber\\
    \tilde{\mu}_{3}&=&\mu _{3}\left(c_{\beta }^2-s_{\beta }^2 \right)+2 \left(\mu _{2}-\mu _{1}\right)c_{\beta } s_{\beta }.
\end{eqnarray}
The quartic couplings that relate only to the two Higgs doublets are given by
\begin{eqnarray}\label{eq:qcHG1}
    \tilde{\lambda}_1&=&\lambda _1c_{\beta }^4+\lambda _2s_{\beta }^4+2\left( \lambda _3+ \lambda _4+{\rm Re}\lambda _5\right)
    c_{\beta }^2 s_{\beta }^2+4{\rm Re}\lambda _6 c_{\beta }^3s_{\beta }+4{\rm Re}\lambda _7 c_{\beta }s_{\beta }^3,\nonumber\\
    \tilde{\lambda}_2&=&\lambda _1s_{\beta }^4+\lambda _2c_{\beta }^4+2\left( \lambda _3+ \lambda _4+{\rm Re}\lambda _5\right)
    c_{\beta }^2 s_{\beta }^2-4{\rm Re}\lambda _6 c_{\beta }s_{\beta }^3-4{\rm Re}\lambda _7 c_{\beta }^3s_{\beta },\nonumber\\
    \tilde{\lambda}_3&=&\frac{1}{4}s_{2\beta}^2\left[ \lambda _1+\lambda _2-2\left( \lambda _3+\lambda _4+{\rm Re}\lambda _5 \right) \right]+\lambda_3
    -\left( {\rm Re}\lambda _6-{\rm Re}\lambda _7 \right)c_{2\beta}s_{2\beta},\nonumber\\
    \tilde{\lambda}_4&=&\frac{1}{4}s_{2\beta}^2\left[ \lambda _1+\lambda _2-2\left( \lambda _3+\lambda _4+{\rm Re}\lambda _5 \right) \right]+\lambda_4
    -\left( {\rm Re}\lambda _6-{\rm Re}\lambda _7 \right)c_{2\beta}s_{2\beta},\nonumber\\
    \tilde{\lambda}_5&=&\left[\lambda _1+\lambda _2-2\left(\lambda _3+\lambda _4\right)\right] c_{\beta}^2 s_{\beta }^2+\lambda _5 c_{\beta}^4+\lambda _5^* s_{\beta}^4
    -2\left(\lambda _6-\lambda _7\right) c_{\beta }^3s_{\beta }\nonumber\\& &+2 \left(\lambda _6^*-\lambda _7^*\right)c_{\beta } s_{\beta}^3,\nonumber\\
    \tilde{\lambda}_6&=&\left(-\lambda _1+\lambda _3+\lambda _4+\lambda _5^*\right)c_{\beta }^3 s_{\beta}+\left(\lambda _2-\lambda _3-\lambda _4-\lambda _5\right)c_{\beta } s_{\beta}^3
    +\lambda _6^* c_{\beta}^4-\lambda _7 s_{\beta }^4\nonumber\\& &-\left(\lambda _6^*-2\lambda _7^*+2 \lambda _6-\lambda _7\right)c_{\beta }^2 s_{\beta}^2,\nonumber\\
    \tilde{\lambda}_7&=&\left(-\lambda _1+\lambda _3+\lambda _4+\lambda _5\right)c_{\beta } s_{\beta}^3-\left(-\lambda _2+\lambda _3+\lambda _4+\lambda _5^*\right)c_{\beta }^3 s_{\beta}
    -\lambda _6 s_{\beta }^4+\lambda _7^* c_{\beta}^4\nonumber\\& &+\left(2\lambda _6^*-\lambda _7^*+\lambda _6-2 \lambda _7\right)c_{\beta }^2 s_{\beta}^2.
\end{eqnarray}
Finally, the quartic couplings involving the complex Higgs triplet are give by
\begin{eqnarray}\label{eq:qcHG2}
    \tilde{\lambda}_8&=&\lambda _8c_{\beta }^2+\lambda _9s_{\beta }^2+2{\rm Re}\lambda _{10} c_{\beta }s_{\beta },\nonumber\\
    \tilde{\lambda}_9&=&\lambda _8s_{\beta }^2+\lambda _9c_{\beta }^2-2{\rm Re}\lambda _{10} c_{\beta }s_{\beta },\nonumber\\
    \tilde{\lambda}_{10}&=&\left(-\lambda _8+\lambda _9\right) c_{\beta }s_{\beta }+\lambda _{10} c_{\beta}^2-\lambda _{10}^* s_{\beta}^2,\nonumber\\
    \tilde{\lambda}_{11}&=&\lambda _{11}c_{\beta }^2+\lambda _{12}s_{\beta }^2+2{\rm Re\lambda_{13}}s_{2\beta},\nonumber\\
    \tilde{\lambda}_{12}&=&\lambda _{11}s_{\beta }^2+\lambda _{12}c_{\beta }^2-2{\rm Re\lambda_{13}}s_{2\beta},\nonumber\\
    \tilde{\lambda}_{13}&=&\left(-\lambda _{11}+\lambda _{12}\right)c_{\beta } s_{\beta }+\lambda _{13}c_{\beta }^2-\lambda _{13}^*s_{\beta}^2,\nonumber\\
    \tilde{\lambda}_{\Delta 1}&=&\lambda_{\Delta 1},~~\tilde{\lambda}_{\Delta 2}=\lambda_{\Delta 2}.
\end{eqnarray}

Note that under the $U(1)$ transformation $H_1 \rightarrow e^{i \chi} H_1$ and $H_2 \rightarrow e^{-i \chi} H_2$,
the scalar potential remains unchanged if the complex parameters of the scalar potential in the Higgs
basis are transformed by the corresponding phase rotation~\cite{Davidson:2005cw}:
\begin{equation}\label{eq:replace2}
    \tilde{\lambda}_5 \rightarrow e^{4 i \chi} \tilde{\lambda}_5\text{ and } 
    \tilde{m}_{12}^2, \tilde{\lambda}_6, \tilde{\lambda}_7, \tilde{\lambda}_{10}, \tilde{\lambda}_{13} \rightarrow 
    e^{2 i \chi}\left\{\tilde{m}_{12}^2, \tilde{\lambda}_6, \tilde{\lambda}_7, \tilde{\lambda}_{10}, \tilde{\lambda}_{13}\right\}.
\end{equation}
Therefore, beginning with the Eqs.~\eqref{eq:dimHG}-\eqref{eq:qcHG2} in the phase $\{\xi=0,\chi=0\}$, we can obtain the relations between the two 
sets of parameters with arbitrary phases $\{\xi,\chi\}$ by applying the replacements~\eqref{eq:replace1} and \eqref{eq:replace2} 
to the Eqs.~\eqref{eq:dimHG}-\eqref{eq:qcHG2}.
Finally, The inversion of Eqs.~\eqref{eq:dimHG}-\eqref{eq:qcHG2} can be obtained by making the
replacements $\tilde{m}_{12}^2\to m_{12}^2$, $\tilde{\lambda}_i\to \lambda_i$, and $\beta\to -\beta$.

\section{Mass matrices in the complex triplet extension of 2HDM}
\label{apd:Mmatrix}

Here we provide some of the mass matrices elements for the extension of 2HDM with an additional complex Higgs triplet. 
In this appendix we express the neutral scalar in the triplet~\eqref{eq:ctriplet} as $\delta^0= \frac{1}{\sqrt{2}}(d^0+i\eta^0)$.
The two doublet scalars in the Higgs basis can be expressed in components as follows, 
\begin{equation}
    \label{eq:Higgsbasis}
    H_{1}=\left(\begin{array}{c}
    H_1^{+} \\
    \frac{1}{\sqrt{2}}\left(v+h_{1}^0+i A_1\right)
    \end{array}\right),~~{\rm and}~~
    H_{2}=\left(\begin{array}{c}
    H^{+} \\
    \frac{1}{\sqrt{2}}\left(h_{2}^0+i A_2 \right),
    \end{array}\right)
\end{equation}
where $h_{1}^0$ and $h_{2}^0$ are CP-even neutral Higgs bosons, $A_2$ and $H^+$ are the physical neutral pseudoscalar and the charged scalar, 
respectively, while $H_1^{\pm}$and $A_1$ are the Goldstone bosons associated with the $W^{\pm}$ and $Z$ gauge bosons.

The mass matrix elements for the $CP$-even components of the neutral scalars in the model are given by
\begin{eqnarray}
    m_{h_1^0h_1^0}^2&=&\frac{3\tilde{\lambda }_1 v^2}{2}+\frac{v_{\Delta}^2 \tilde{\lambda }_8}{2}+\frac{v_{\Delta}^2\tilde{\lambda }_{11}}{2}-\sqrt{2}
    v_{\Delta} \tilde{\mu} _{1}+\tilde{m}_1^2,\nonumber\\
    m_{h_2^0h_2^0}^2&=&\frac{\tilde{\lambda }_3v^2}{2}+\frac{\tilde{\lambda }_4 v^2}{2}+\frac{1}{4} v^2{\rm Re}\tilde{\lambda }_5+
    \frac{v_{\Delta}^2\tilde{\lambda }_9}{2}+\frac{v_{\Delta}^2\tilde{\lambda }_{12}}{2}-\sqrt{2}v_{\Delta} \tilde{\mu} _{2}+\tilde{m}_2^2,\nonumber\\
    m_{d^0d^0}^2&=&3 v_{\Delta}^2 \tilde{\lambda }_{\Delta 1}+3 v_{\Delta}^2\tilde{\lambda }_{\Delta 2}+\tilde{m}_{\Delta}^2+\frac{\tilde{\lambda }_8v^2}{2}
    +\frac{\tilde{\lambda }_{11} v^2}{2},\nonumber\\
    m_{h_1^0h_2^0}^2&=&\frac{3}{4} v^2{\rm Re}\tilde{\lambda }_6+\frac{1}{4} v_{\Delta}^2{\rm Re}\tilde{\lambda }_{10}+\frac{1}{4}v_{\Delta}^2{\rm Re}\tilde{\lambda }_{13}-\frac{v_{\Delta}{\rm Re} \tilde{\mu} _{3}}{2
    \sqrt{2}}-\frac{{\rm Re}\tilde{m}_{12}^{2}}{2},\nonumber\\
    m_{h_1^0d^0}^2&=&v_{\Delta} \tilde{\lambda }_8v+v_{\Delta} \tilde{\lambda }_{11} v-\sqrt{2}\tilde{\mu} _{1} v,\nonumber\\
    m_{h_2^0d^0}^2&=&\frac{1}{2} v_{\Delta}v {\rm Re}\tilde{\lambda }_{10}+\frac{1}{2}v_{\Delta} v {\rm Re}\tilde{\lambda }_{13}
    -\frac{v {\rm Re}\tilde{\mu} _{3}}{2\sqrt{2}}\,,
\end{eqnarray}
while the mass matrix elements for the $CP$-odd components are shown as follows % of the neutral scalars are given by
\begin{eqnarray}
    m_{A_1^0A_1^0}^2&=&\frac{\tilde{\lambda }_1 v^2}{2}+\frac{v_{\Delta}^2 \tilde{\lambda }_8}{2}+\frac{v_{\Delta}^2\tilde{\lambda }_{11}}{2}
    +\sqrt{2}v_{\Delta} \tilde{\mu} _{1}+\tilde{m}_1^2,\nonumber\\
    m_{A_2^0A_2^0}^2&=&\frac{\tilde{\lambda }_3v^2}{2}+\frac{\tilde{\lambda }_4 v^2}{2}-\frac{1}{4} v^2{\rm Re}\tilde{\lambda }_5+\frac{v_{\Delta}^2\tilde{\lambda }_9}{2}+\frac{v_{\Delta}^2\tilde{\lambda }_{12}}{2}
    +\sqrt{2}v_{\Delta} \tilde{\mu} _{2}+\tilde{m}_2^2,\nonumber\\
    m_{\eta^0\eta^0}^2&=&v_{\Delta}^2 \tilde{\lambda }_{\Delta 1}+v_{\Delta}^2\tilde{\lambda }_{\Delta 2}+\tilde{m}_{\Delta}^2
    +\frac{\tilde{\lambda }_8v^2}{2}+\frac{\tilde{\lambda }_{11} v^2}{2},\nonumber\\
    m_{A_1^0A_2^0}^2&=&\frac{v_{\Delta}^2{\rm Re}\tilde{\lambda }_{10}}{4}+\frac{v_{\Delta}^2{\rm Re}\tilde{\lambda }_{13}}{4}+
    \frac{v_{\Delta}{\rm Re}\tilde{\mu} _{3}}{2\sqrt{2}}-\frac{{\rm Re}\tilde{m}_{12}^2}{2}+\frac{{\rm Re}\tilde{\lambda }_6v^2}{4},\nonumber\\
    m_{A_1^0\eta^0}^2&=&-\sqrt{2} \tilde{\mu} _{1} v,\nonumber\\
    m_{A_2^0\eta^0}^2&=&-\frac{{\rm Re}\tilde{\mu} _{3}v}{2 \sqrt{2}}.
\end{eqnarray}

The mass matrix elements for the singly-charged scalars are
\begin{eqnarray}\label{eq:SCmass}
    m_{H_1^+H_1^-}^2&=&\frac{\tilde{\lambda }_1 v^2}{2}+\frac{v_{\Delta}^2 \tilde{\lambda }_8}{2}+\tilde{m}_1^2,\nonumber\\
    m_{H_2^+H_2^-}^2&=&\frac{\tilde{\lambda }_3 v^2}{2}+\frac{v_{\Delta}^2 \tilde{\lambda }_9}{2}+\tilde{m}_2^2,\nonumber\\
    m_{\delta^+\delta^-}^2&=&\frac{\tilde{\lambda }_8v^2}{2}+\frac{\tilde{\lambda }_{11} v^2}{4}+
    v_{\Delta}^2 \tilde{\lambda }_{\Delta 1}+v_{\Delta}^2\tilde{\lambda }_{\Delta 2}+\tilde{m}_{\Delta}^2,\nonumber\\
    m_{H_1^+H_2^-}^2&=&\frac{1}{2} v^2 \tilde{\lambda }_6^*+\frac{1}{2} v_{\Delta}^2\tilde{\lambda }_{10}^*-\tilde{m}_{12}^{2*}=m_{H_1^-H_2^+}^{2*},\nonumber\\
    m_{H_1^+\delta^-}^2&=&\frac{v_{\Delta} \tilde{\lambda }_{11}v}{2 \sqrt{2}}-\tilde{\mu} _{1} v=m_{H_1^-\delta^+}^{2*},\nonumber\\
    m_{H_2^+\delta^-}^2&=&\frac{v_{\Delta} \tilde{\lambda }_{13}v}{2\sqrt{2}}-\frac{\tilde{\mu} _{3}v}{2}=m_{H_2^-\delta^+}^{2*}.
\end{eqnarray}

Since there is only one doubly-charged scalar in the model, so there is not any mixing and its mass is simply given by
\begin{equation}\label{eq:DCmass}
    m_{\delta^{++}\delta^{--}}^2=v_{\Delta}^2 \tilde{\lambda}_{\Delta 1}+\tilde{m}_{\Delta}^2+\frac{\tilde{\lambda }_8 v^2}{2}
\end{equation}

The parameters with tilde denote those in the Higgs basis, with the transformation relation to the parameters in the generic basis 
given in Appendix~\ref{apd:HBcouplings}. Due to $CP$-violating effects, there may exist mixings between the $CP$-even and $CP$-odd components 
in the neutral scalars. Since we do not use them in our work, we do not provide their explicit formulae here.

\section{Trilinear couplings in the complex triplet extension of 2HDM} 
\label{apd:tricoupling}

The trilinear couplings between a neutral scalar and a pair of charged scalars in the triplet extension of the 2HDM are summarized as follows:
\begin{eqnarray}\label{eq:tricp}
    \lambda_{h_1^0H_2^+H_2^-}=\tilde{\lambda}_3&,&~~
    \lambda_{h_2^0H_2^+H_2^-}={\rm Re}\tilde{\lambda}_7,\nonumber\\
    \lambda_{h_1^0\delta^+\delta^-}=\tilde{\lambda} _8+\frac{1}{2}\tilde{\lambda} _{11}&,&~~
    \lambda_{h_2^0\delta^+\delta^-}={\rm Re}\tilde{\lambda}_{10}+\frac{1}{2}{\rm Re}\tilde{\lambda}_{13},\nonumber\\
    \lambda_{h_1^0\delta^{++}\delta^{--}}=\tilde{\lambda} _8&,&~~
    \lambda_{h_2^0\delta^{++}\delta^{--}}={\rm Re}\tilde{\lambda}_{10}.
\end{eqnarray}
Note that in the aligned and decoupling limits, we have $h_1^0\equiv h$ and $h_2^0\equiv H$.


\begin{thebibliography}{200}

%====================================================================== 
%\cite{ATLAS:2012yve}
\bibitem{ATLAS:2012yve}
G.~Aad \textit{et al.} [ATLAS],
``Observation of a new particle in the search for the Standard Model Higgs boson with the ATLAS detector at the LHC,''
Phys. Lett. B \textbf{716}, 1-29 (2012)
doi:10.1016/j.physletb.2012.08.020
[arXiv:1207.7214 [hep-ex]].
%13646 citations counted in INSPIRE as of 17 Oct 2022

%\cite{CMS:2012qbp}
\bibitem{CMS:2012qbp}
S.~Chatrchyan \textit{et al.} [CMS],
``Observation of a New Boson at a Mass of 125 GeV with the CMS Experiment at the LHC,''
Phys. Lett. B \textbf{716}, 30-61 (2012)
doi:10.1016/j.physletb.2012.08.021
[arXiv:1207.7235 [hep-ex]].
%13272 citations counted in INSPIRE as of 17 Oct 2022

%\cite{Formaggio:2021nfz}
\bibitem{Formaggio:2021nfz}
J.~A.~Formaggio, A.~L.~C.~de Gouv\^ea and R.~G.~H.~Robertson,
``Direct Measurements of Neutrino Mass,''
Phys. Rept. \textbf{914}, 1-54 (2021)
doi:10.1016/j.physrep.2021.02.002
[arXiv:2102.00594 [nucl-ex]].
%39 citations counted in INSPIRE as of 18 Oct 2022

%\cite{Planck:2015fie}
\bibitem{Planck:2015fie}
P.~A.~R.~Ade \textit{et al.} [Planck],
``Planck 2015 results. XIII. Cosmological parameters,''
Astron. Astrophys. \textbf{594}, A13 (2016)
doi:10.1051/0004-6361/201525830
[arXiv:1502.01589 [astro-ph.CO]].
%11099 citations counted in INSPIRE as of 18 Oct 2022

%\cite{Kuzmin:1985mm}
\bibitem{Kuzmin:1985mm}
V.~A.~Kuzmin, V.~A.~Rubakov and M.~E.~Shaposhnikov,
``On the Anomalous Electroweak Baryon Number Nonconservation in the Early Universe,''
Phys. Lett. B \textbf{155}, 36 (1985)
doi:10.1016/0370-2693(85)91028-7
%3200 citations counted in INSPIRE as of 18 Oct 2022

%\cite{Degrassi:2021uik}
\bibitem{Degrassi:2021uik}
G.~Degrassi, B.~Di Micco, P.~P.~Giardino and E.~Rossi,
``Higgs boson self-coupling constraints from single Higgs, double Higgs and Electroweak measurements,''
Phys. Lett. B \textbf{817}, 136307 (2021)
doi:10.1016/j.physletb.2021.136307
[arXiv:2102.07651 [hep-ph]].
%7 citations counted in INSPIRE as of 18 Oct 2022

%\cite{McCullough:2013rea}
\bibitem{McCullough:2013rea}
M.~McCullough,
``An Indirect Model-Dependent Probe of the Higgs Self-Coupling,''
Phys. Rev. D \textbf{90}, no.1, 015001 (2014)
[erratum: Phys. Rev. D \textbf{92}, no.3, 039903 (2015)]
doi:10.1103/PhysRevD.90.015001
[arXiv:1312.3322 [hep-ph]].
%146 citations counted in INSPIRE as of 18 Oct 2022

%\cite{Cao:2015oxx}
\bibitem{Cao:2015oxx}
Q.~H.~Cao, Y.~Liu and B.~Yan,
``Measuring trilinear Higgs coupling in WHH and ZHH productions at the high-luminosity LHC,''
Phys. Rev. D \textbf{95}, no.7, 073006 (2017)
doi:10.1103/PhysRevD.95.073006
[arXiv:1511.03311 [hep-ph]].
%50 citations counted in INSPIRE as of 18 Oct 2022

%\cite{Bizon:2016wgr}
\bibitem{Bizon:2016wgr}
W.~Bizon, M.~Gorbahn, U.~Haisch and G.~Zanderighi,
``Constraints on the trilinear Higgs coupling from vector boson fusion and associated Higgs production at the LHC,''
JHEP \textbf{07}, 083 (2017)
doi:10.1007/JHEP07(2017)083
[arXiv:1610.05771 [hep-ph]].
%81 citations counted in INSPIRE as of 18 Oct 2022

%\cite{deBlas:2016ojx}
\bibitem{deBlas:2016ojx}
J.~de Blas, M.~Ciuchini, E.~Franco, S.~Mishima, M.~Pierini, L.~Reina and L.~Silvestrini,
``Electroweak precision observables and Higgs-boson signal strengths in the Standard Model and beyond: present and future,''
JHEP \textbf{12}, 135 (2016)
doi:10.1007/JHEP12(2016)135
[arXiv:1608.01509 [hep-ph]].
%163 citations counted in INSPIRE as of 18 Oct 2022

%======================================================================
%\cite{Gell-Mann:1969cuq,Weinberg:1971fb,Lee:1977yc,Lee:1977eg}
\bibitem{Gell-Mann:1969cuq}
M.~Gell-Mann, M.~L.~Goldberger, N.~M.~Kroll and F.~E.~Low,
``Amelioration of divergence difficulties in the theory of weak interactions,''
Phys. Rev. \textbf{179}, 1518-1527 (1969)
doi:10.1103/PhysRev.179.1518
%125 citations counted in INSPIRE as of 23 Jul 2022

%\cite{Weinberg:1971fb}
\bibitem{Weinberg:1971fb}
S.~Weinberg,
``Physical Processes in a Convergent Theory of the Weak and Electromagnetic Interactions,''
Phys. Rev. Lett. \textbf{27}, 1688-1691 (1971)
doi:10.1103/PhysRevLett.27.1688
%727 citations counted in INSPIRE as of 23 Jul 2022

%\cite{Lee:1977yc}
\bibitem{Lee:1977yc}
B.~W.~Lee, C.~Quigg and H.~B.~Thacker,
``The Strength of Weak Interactions at Very High-Energies and the Higgs Boson Mass,''
Phys. Rev. Lett. \textbf{38}, 883-885 (1977)
doi:10.1103/PhysRevLett.38.883
%843 citations counted in INSPIRE as of 07 Oct 2022

%\cite{Lee:1977eg}
\bibitem{Lee:1977eg}
B.~W.~Lee, C.~Quigg and H.~B.~Thacker,
``Weak Interactions at Very High-Energies: The Role of the Higgs Boson Mass,''
Phys. Rev. D \textbf{16}, 1519 (1977)
doi:10.1103/PhysRevD.16.1519
%2321 citations counted in INSPIRE as of 07 Oct 2022

%======================================================================
%\cite{Cabibbo:1979ay}
\bibitem{Cabibbo:1979ay}
N.~Cabibbo, L.~Maiani, G.~Parisi and R.~Petronzio,
``Bounds on the Fermions and Higgs Boson Masses in Grand Unified Theories,''
Nucl. Phys. B \textbf{158}, 295-305 (1979)
doi:10.1016/0550-3213(79)90167-6
%838 citations counted in INSPIRE as of 16 Oct 2022

%\cite{Lindner:1985uk}
\bibitem{Lindner:1985uk}
M.~Lindner,
``Implications of Triviality for the Standard Model,''
Z. Phys. C \textbf{31}, 295 (1986)
doi:10.1007/BF01479540
%501 citations counted in INSPIRE as of 16 Oct 2022

%======================================================================

%\cite{Branco:2011iw}
\bibitem{Branco:2011iw}
G.~C.~Branco, P.~M.~Ferreira, L.~Lavoura, M.~N.~Rebelo, M.~Sher and J.~P.~Silva,
``Theory and phenomenology of two-Higgs-doublet models,''
Phys. Rept. \textbf{516}, 1-102 (2012)
doi:10.1016/j.physrep.2012.02.002
[arXiv:1106.0034 [hep-ph]].
%2427 citations counted in INSPIRE as of 18 Oct 2022

%\cite{Wang:2022yhm}
\bibitem{Wang:2022yhm}
L.~Wang, J.~M.~Yang and Y.~Zhang,
``Two-Higgs-doublet models in light of current experiments: a brief review,''
Commun. Theor. Phys. \textbf{74}, no.9, 097202 (2022)
doi:10.1088/1572-9494/ac7fe9
[arXiv:2203.07244 [hep-ph]].
%6 citations counted in INSPIRE as of 18 Oct 2022

%\cite{Huffel:1980sk}
\bibitem{Huffel:1980sk}
H.~Huffel and G.~Pocsik,
``Unitarity Bounds on Higgs Boson Masses in the {Weinberg-Salam} Model With Two Higgs Doublets,''
Z. Phys. C \textbf{8}, 13 (1981)
doi:10.1007/BF01429824
%144 citations counted in INSPIRE as of 07 Oct 2022

%\cite{Maalampi:1991fb}
\bibitem{Maalampi:1991fb}
J.~Maalampi, J.~Sirkka and I.~Vilja,
``Tree level unitarity and triviality bounds for two Higgs models,''
Phys. Lett. B \textbf{265}, 371-376 (1991)
doi:10.1016/0370-2693(91)90068-2
%107 citations counted in INSPIRE as of 07 Oct 2022

%\cite{Kanemura:1993hm}
\bibitem{Kanemura:1993hm}
S.~Kanemura, T.~Kubota and E.~Takasugi,
``Lee-Quigg-Thacker bounds for Higgs boson masses in a two doublet model,''
Phys. Lett. B \textbf{313}, 155-160 (1993)
doi:10.1016/0370-2693(93)91205-2
[arXiv:hep-ph/9303263 [hep-ph]].
%408 citations counted in INSPIRE as of 09 Feb 2023

%\cite{Akeroyd:2000wc}
\bibitem{Akeroyd:2000wc}
A.~G.~Akeroyd, A.~Arhrib and E.~M.~Naimi,
``Note on tree level unitarity in the general two Higgs doublet model,''
Phys. Lett. B \textbf{490}, 119-124 (2000)
doi:10.1016/S0370-2693(00)00962-X
[arXiv:hep-ph/0006035 [hep-ph]].
%387 citations counted in INSPIRE as of 07 Oct 2022

%\cite{Arhrib:2000is}
\bibitem{Arhrib:2000is}
A.~Arhrib,
``Unitarity constraints on scalar parameters of the standard and two Higgs doublets model,''
[arXiv:hep-ph/0012353 [hep-ph]].
%105 citations counted in INSPIRE as of 07 Oct 2022

%======================================================================

%\cite{Ginzburg:2005dt}
\bibitem{Ginzburg:2005dt}
I.~F.~Ginzburg and I.~P.~Ivanov,
``Tree-level unitarity constraints in the most general 2HDM,''
Phys. Rev. D \textbf{72}, 115010 (2005)
doi:10.1103/PhysRevD.72.115010
[arXiv:hep-ph/0508020 [hep-ph]].
%276 citations counted in INSPIRE as of 07 Oct 2022

%\cite{Kanemura:2015ska}
\bibitem{Kanemura:2015ska}
S.~Kanemura and K.~Yagyu,
``Unitarity bound in the most general two Higgs doublet model,''
Phys. Lett. B \textbf{751}, 289-296 (2015)
doi:10.1016/j.physletb.2015.10.047
[arXiv:1509.06060 [hep-ph]].
%89 citations counted in INSPIRE as of 07 Oct 2022

%======================================================================
%\cite{Georgi:1985nv}
\bibitem{Georgi:1985nv}
H.~Georgi and M.~Machacek,
``DOUBLY CHARGED HIGGS BOSONS,''
Nucl. Phys. B \textbf{262}, 463-477 (1985)
doi:10.1016/0550-3213(85)90325-6
%504 citations counted in INSPIRE as of 26 Nov 2022

%\cite{Aoki:2007ah}
\bibitem{Aoki:2007ah}
M.~Aoki and S.~Kanemura,
``Unitarity bounds in the Higgs model including triplet fields with custodial symmetry,''
Phys. Rev. D \textbf{77}, no.9, 095009 (2008)
[erratum: Phys. Rev. D \textbf{89}, no.5, 059902 (2014)]
doi:10.1103/PhysRevD.77.095009
[arXiv:0712.4053 [hep-ph]].
%94 citations counted in INSPIRE as of 07 Oct 2022

%\cite{Konetschny:1977bn,Cheng:1980qt,Magg:1980ut,Schechter:1980gr,Lazarides:1980nt,Mohapatra:1980yp}
\bibitem{Konetschny:1977bn}
W.~Konetschny and W.~Kummer,
``Nonconservation of Total Lepton Number with Scalar Bosons,''
Phys. Lett. B \textbf{70}, 433-435 (1977)
doi:10.1016/0370-2693(77)90407-5
%421 citations counted in INSPIRE as of 26 Nov 2022

%\cite{Cheng:1980qt}
\bibitem{Cheng:1980qt}
T.~P.~Cheng and L.~F.~Li,
``Neutrino Masses, Mixings and Oscillations in SU(2) x U(1) Models of Electroweak Interactions,''
Phys. Rev. D \textbf{22}, 2860 (1980)
doi:10.1103/PhysRevD.22.2860
%1202 citations counted in INSPIRE as of 26 Nov 2022

%\cite{Magg:1980ut}
\bibitem{Magg:1980ut}
M.~Magg and C.~Wetterich,
``Neutrino Mass Problem and Gauge Hierarchy,''
Phys. Lett. B \textbf{94}, 61-64 (1980)
doi:10.1016/0370-2693(80)90825-4
%1128 citations counted in INSPIRE as of 26 Nov 2022

%\cite{Schechter:1980gr}
\bibitem{Schechter:1980gr}
J.~Schechter and J.~W.~F.~Valle,
``Neutrino Masses in SU(2) x U(1) Theories,''
Phys. Rev. D \textbf{22}, 2227 (1980)
doi:10.1103/PhysRevD.22.2227
%3355 citations counted in INSPIRE as of 26 Nov 2022

%\cite{Lazarides:1980nt}
\bibitem{Lazarides:1980nt}
G.~Lazarides, Q.~Shafi and C.~Wetterich,
``Proton Lifetime and Fermion Masses in an SO(10) Model,''
Nucl. Phys. B \textbf{181}, 287-300 (1981)
doi:10.1016/0550-3213(81)90354-0
%1601 citations counted in INSPIRE as of 26 Nov 2022

%\cite{Mohapatra:1980yp}
\bibitem{Mohapatra:1980yp}
R.~N.~Mohapatra and G.~Senjanovic,
``Neutrino Masses and Mixings in Gauge Models with Spontaneous Parity Violation,''
Phys. Rev. D \textbf{23}, 165 (1981)
doi:10.1103/PhysRevD.23.165
%2862 citations counted in INSPIRE as of 26 Nov 2022

%\cite{Arhrib:2011uy}
\bibitem{Arhrib:2011uy}
A.~Arhrib, R.~Benbrik, M.~Chabab, G.~Moultaka, M.~C.~Peyranere, L.~Rahili and J.~Ramadan,
``The Higgs Potential in the Type II Seesaw Model,''
Phys. Rev. D \textbf{84}, 095005 (2011)
doi:10.1103/PhysRevD.84.095005
[arXiv:1105.1925 [hep-ph]].
%196 citations counted in INSPIRE as of 07 Oct 2022

%\cite{Khan:2016sxm}
\bibitem{Khan:2016sxm}
N.~Khan,
``Exploring the hyperchargeless Higgs triplet model up to the Planck scale,''
Eur. Phys. J. C \textbf{78}, no.4, 341 (2018)
doi:10.1140/epjc/s10052-018-5766-4
[arXiv:1610.03178 [hep-ph]].
%49 citations counted in INSPIRE as of 07 Oct 2022

%\cite{Ouazghour:2018mld}
\bibitem{Ouazghour:2018mld}
B.~A.~Ouazghour, A.~Arhrib, R.~Benbrik, M.~Chabab and L.~Rahili,
``Theory and phenomenology of a two-Higgs-doublet type-II seesaw model at the LHC run 2,''
Phys. Rev. D \textbf{100}, no.3, 035031 (2019)
doi:10.1103/PhysRevD.100.035031
[arXiv:1812.07719 [hep-ph]].
%7 citations counted in INSPIRE as of 07 Oct 2022

%\cite{Cao:2013wqa}
\bibitem{Cao:2013wqa}
J.~Cao, P.~Wan, J.~M.~Yang and J.~Zhu,
``The SM extension with color-octet scalars: diphoton enhancement and global fit of LHC Higgs data,''
JHEP \textbf{08}, 009 (2013)
doi:10.1007/JHEP08(2013)009
[arXiv:1303.2426 [hep-ph]].
%52 citations counted in INSPIRE as of 09 Feb 2023

%\cite{Chang:2019vez,Abu-Ajamieh:2020yqi,Abu-Ajamieh:2021egq,Abu-Ajamieh:2022ppp}
\bibitem{Chang:2019vez}
S.~Chang and M.~A.~Luty,
``The Higgs Trilinear Coupling and the Scale of New Physics,''
JHEP \textbf{03}, 140 (2020)
doi:10.1007/JHEP03(2020)140
[arXiv:1902.05556 [hep-ph]].
%45 citations counted in INSPIRE as of 21 Mar 2023

%\cite{Abu-Ajamieh:2020yqi}
% \bibitem{Abu-Ajamieh:2020yqi}
% F.~Abu-Ajamieh, S.~Chang, M.~Chen and M.~A.~Luty,
% %``Higgs coupling measurements and the scale of new physics,''
% JHEP \textbf{07}, 056 (2021)
% doi:10.1007/JHEP07(2021)056
% [arXiv:2009.11293 [hep-ph]].
% %15 citations counted in INSPIRE as of 21 Mar 2023

%\cite{Abu-Ajamieh:2021egq}
\bibitem{Abu-Ajamieh:2021egq}
F.~Abu-Ajamieh,
``The scale of new physics from the Higgs couplings to \ensuremath{\gamma}\ensuremath{\gamma} and \ensuremath{\gamma}Z,''
JHEP \textbf{06}, 091 (2022)
doi:10.1007/JHEP06(2022)091
[arXiv:2112.13529 [hep-ph]].
%3 citations counted in INSPIRE as of 21 Mar 2023

% %\cite{Abu-Ajamieh:2022ppp}
% \bibitem{Abu-Ajamieh:2022ppp}
% F.~Abu-Ajamieh,
% %``The scale of new physics from the Higgs couplings to gg,''
% Phys. Lett. B \textbf{833}, 137389 (2022)
% doi:10.1016/j.physletb.2022.137389
% [arXiv:2203.07410 [hep-ph]].
%2 citations counted in INSPIRE as of 21 Mar 2023

%======================================================================
%\cite{Cornwall:1973tb}
\bibitem{Cornwall:1973tb}
J.~M.~Cornwall, D.~N.~Levin and G.~Tiktopoulos,
``Uniqueness of spontaneously broken gauge theories,''
Phys. Rev. Lett. \textbf{30}, 1268-1270 (1973)
[erratum: Phys. Rev. Lett. \textbf{31}, 572 (1973)]
doi:10.1103/PhysRevLett.30.1268
%425 citations counted in INSPIRE as of 08 Oct 2022

%\cite{Cornwall:1974km}
\bibitem{Cornwall:1974km}
J.~M.~Cornwall, D.~N.~Levin and G.~Tiktopoulos,
``Derivation of Gauge Invariance from High-Energy Unitarity Bounds on the s Matrix,''
Phys. Rev. D \textbf{10}, 1145 (1974)
[erratum: Phys. Rev. D \textbf{11}, 972 (1975)]
doi:10.1103/PhysRevD.10.1145
%1326 citations counted in INSPIRE as of 08 Oct 2022

%\cite{Yao:1988aj}
\bibitem{Yao:1988aj}
Y.~P.~Yao and C.~P.~Yuan,
``Modification of the Equivalence Theorem Due to Loop Corrections,''
Phys. Rev. D \textbf{38} (1988), 2237
doi:10.1103/PhysRevD.38.2237
%190 citations counted in INSPIRE as of 19 Nov 2022

%\cite{Veltman:1989ud}
\bibitem{Veltman:1989ud}
H.~G.~J.~Veltman,
``The Equivalence Theorem,''
Phys. Rev. D \textbf{41} (1990), 2294
doi:10.1103/PhysRevD.41.2294
%211 citations counted in INSPIRE as of 19 Nov 2022

%\cite{He:1992nga}
\bibitem{He:1992nga}
H.~J.~He, Y.~P.~Kuang and X.~y.~Li,
``On the precise formulation of equivalence theorem,''
Phys. Rev. Lett. \textbf{69} (1992), 2619-2622
doi:10.1103/PhysRevLett.69.2619
%178 citations counted in INSPIRE as of 19 Nov 2022

%======================================================================
%\cite{FileviezPerez:2008bj}
\bibitem{FileviezPerez:2008bj}
P.~Fileviez Perez, H.~H.~Patel, M.~J.~Ramsey-Musolf and K.~Wang,
``Triplet Scalars and Dark Matter at the LHC,''
Phys. Rev. D \textbf{79}, 055024 (2009)
doi:10.1103/PhysRevD.79.055024
[arXiv:0811.3957 [hep-ph]].
%132 citations counted in INSPIRE as of 11 Oct 2022

%\cite{YaserAyazi:2014jby}
\bibitem{YaserAyazi:2014jby}
S.~Yaser Ayazi and S.~M.~Firouzabadi,
``Constraining Inert Triplet Dark Matter by the LHC and FermiLAT,''
JCAP \textbf{11}, 005 (2014)
doi:10.1088/1475-7516/2014/11/005
[arXiv:1408.0654 [hep-ph]].
%15 citations counted in INSPIRE as of 11 Oct 2022

%\cite{Chiang:2020rcv}
\bibitem{Chiang:2020rcv}
C.~W.~Chiang, G.~Cottin, Y.~Du, K.~Fuyuto and M.~J.~Ramsey-Musolf,
``Collider Probes of Real Triplet Scalar Dark Matter,''
JHEP \textbf{01}, 198 (2021)
doi:10.1007/JHEP01(2021)198
[arXiv:2003.07867 [hep-ph]].
%33 citations counted in INSPIRE as of 11 Oct 2022

%\cite{Wang:2013jba}
\bibitem{Wang:2013jba}
L.~Wang and X.~F.~Han,
``LHC diphoton and Z+photon Higgs signals in the Higgs triplet model with Y = 0,''
JHEP \textbf{03}, 010 (2014)
doi:10.1007/JHEP03(2014)010
[arXiv:1303.4490 [hep-ph]].
%15 citations counted in INSPIRE as of 11 Oct 2022

%\cite{Bandyopadhyay:2014vma}
\bibitem{Bandyopadhyay:2014vma}
P.~Bandyopadhyay, K.~Huitu and A.~Sabanci Keceli,
``Multi-Lepton Signatures of the Triplet Like Charged Higgs at the LHC,''
JHEP \textbf{05}, 026 (2015)
doi:10.1007/JHEP05(2015)026
[arXiv:1412.7359 [hep-ph]].
%28 citations counted in INSPIRE as of 11 Oct 2022

%\cite{Niemi:2018asa}
\bibitem{Niemi:2018asa}
L.~Niemi, H.~H.~Patel, M.~J.~Ramsey-Musolf, T.~V.~I.~Tenkanen and D.~J.~Weir,
``Electroweak phase transition in the real triplet extension of the SM: Dimensional reduction,''
Phys. Rev. D \textbf{100}, no.3, 035002 (2019)
doi:10.1103/PhysRevD.100.035002
[arXiv:1802.10500 [hep-ph]].
%53 citations counted in INSPIRE as of 11 Oct 2022

%\cite{Bell:2020gug}
\bibitem{Bell:2020gug}
N.~F.~Bell, M.~J.~Dolan, L.~S.~Friedrich, M.~J.~Ramsey-Musolf and R.~R.~Volkas,
``Two-Step Electroweak Symmetry-Breaking: Theory Meets Experiment,''
JHEP \textbf{05}, 050 (2020)
doi:10.1007/JHEP05(2020)050
[arXiv:2001.05335 [hep-ph]].
%26 citations counted in INSPIRE as of 11 Oct 2022

%======================================================================
%\cite{Chen:2021jok}
\bibitem{Chen:2021jok}
C.~H.~Chen, C.~W.~Chiang and T.~Nomura,
``Muon g-2 in a two-Higgs-doublet model with a type-II seesaw mechanism,''
Phys. Rev. D \textbf{104}, no.5, 055011 (2021)
doi:10.1103/PhysRevD.104.055011
[arXiv:2104.03275 [hep-ph]].
%27 citations counted in INSPIRE as of 05 Oct 2022

%\cite{Mohapatra:1999zr}
\bibitem{Mohapatra:1999zr}
R.~N.~Mohapatra, A.~Perez-Lorenzana and C.~A.~de Sousa Pires,
``Type II seesaw and a gauge model for the bimaximal mixing explanation of neutrino puzzles,''
Phys. Lett. B \textbf{474}, 355-360 (2000)
doi:10.1016/S0370-2693(00)00026-5
[arXiv:hep-ph/9911395 [hep-ph]].
%151 citations counted in INSPIRE as of 17 Oct 2022

%\cite{Gu:2006wj}
\bibitem{Gu:2006wj}
P.~H.~Gu, H.~Zhang and S.~Zhou,
``A Minimal Type II Seesaw Model,''
Phys. Rev. D \textbf{74}, 076002 (2006)
doi:10.1103/PhysRevD.74.076002
[arXiv:hep-ph/0606302 [hep-ph]].
%44 citations counted in INSPIRE as of 17 Oct 2022

%\cite{Chao:2007mz}
\bibitem{Chao:2007mz}
W.~Chao, S.~Luo, Z.~z.~Xing and S.~Zhou,
``A Compromise between Neutrino Masses and Collider Signatures in the Type-II Seesaw Model,''
Phys. Rev. D \textbf{77}, 016001 (2008)
doi:10.1103/PhysRevD.77.016001
[arXiv:0709.1069 [hep-ph]].
%39 citations counted in INSPIRE as of 17 Oct 2022

%\cite{FileviezPerez:2008jbu}
\bibitem{FileviezPerez:2008jbu}
P.~Fileviez Perez, T.~Han, G.~y.~Huang, T.~Li and K.~Wang,
``Neutrino Masses and the CERN LHC: Testing Type II Seesaw,''
Phys. Rev. D \textbf{78}, 015018 (2008)
doi:10.1103/PhysRevD.78.015018
[arXiv:0805.3536 [hep-ph]].
%363 citations counted in INSPIRE as of 17 Oct 2022

%\cite{Melfo:2011nx}
\bibitem{Melfo:2011nx}
A.~Melfo, M.~Nemevsek, F.~Nesti, G.~Senjanovic and Y.~Zhang,
``Type II Seesaw at LHC: The Roadmap,''
Phys. Rev. D \textbf{85}, 055018 (2012)
doi:10.1103/PhysRevD.85.055018
[arXiv:1108.4416 [hep-ph]].
%188 citations counted in INSPIRE as of 17 Oct 2022

%\cite{Chen:2013dh}
\bibitem{Chen:2013dh}
C.~S.~Chen, C.~Q.~Geng, D.~Huang and L.~H.~Tsai,
``$h\rightarrow Z\gamma$ in Type-II seesaw neutrino model,''
Phys. Lett. B \textbf{723}, 156-160 (2013)
doi:10.1016/j.physletb.2013.05.007
[arXiv:1302.0502 [hep-ph]].
%28 citations counted in INSPIRE as of 17 Oct 2022

%\cite{Han:2015sca}
\bibitem{Han:2015sca}
Z.~L.~Han, R.~Ding and Y.~Liao,
``LHC phenomenology of the type II seesaw mechanism: Observability of neutral scalars in the nondegenerate case,''
Phys. Rev. D \textbf{92}, no.3, 033014 (2015)
doi:10.1103/PhysRevD.92.033014
[arXiv:1506.08996 [hep-ph]].
%34 citations counted in INSPIRE as of 17 Oct 2022

%\cite{Dev:2018sel}
\bibitem{Dev:2018sel}
P.~S.~B.~Dev, M.~J.~Ramsey-Musolf and Y.~Zhang,
``Doubly-Charged Scalars in the Type-II Seesaw Mechanism: Fundamental Symmetry Tests and High-Energy Searches,''
Phys. Rev. D \textbf{98}, no.5, 055013 (2018)
doi:10.1103/PhysRevD.98.055013
[arXiv:1806.08499 [hep-ph]].
%44 citations counted in INSPIRE as of 17 Oct 2022

%\cite{Du:2018eaw}
\bibitem{Du:2018eaw}
Y.~Du, A.~Dunbrack, M.~J.~Ramsey-Musolf and J.~H.~Yu,
``Type-II Seesaw Scalar Triplet Model at a 100 TeV $pp$ Collider: Discovery and Higgs Portal Coupling Determination,''
JHEP \textbf{01}, 101 (2019)
doi:10.1007/JHEP01(2019)101
[arXiv:1810.09450 [hep-ph]].
%46 citations counted in INSPIRE as of 17 Oct 2022

%\cite{Cheng:2022jyi}
\bibitem{Cheng:2022jyi}
Y.~Cheng, X.~G.~He, Z.~L.~Huang and M.~W.~Li,
``Type-II seesaw triplet scalar effects on neutrino trident scattering,''
Phys. Lett. B \textbf{831}, 137218 (2022)
doi:10.1016/j.physletb.2022.137218
[arXiv:2204.05031 [hep-ph]].
%55 citations counted in INSPIRE as of 17 Oct 2022

%\cite{Ding:2017jdr}
\bibitem{Ding:2017jdr}
R.~Ding, Z.~L.~Han, L.~Feng and B.~Zhu,
``Confronting the DAMPE Excess with the Scotogenic Type-II Seesaw Model,''
Chin. Phys. C \textbf{42}, no.8, 083104 (2018)
doi:10.1088/1674-1137/42/8/083104
[arXiv:1712.02021 [hep-ph]].
%31 citations counted in INSPIRE as of 17 Oct 2022

%\cite{Primulando:2019evb}
\bibitem{Primulando:2019evb}
R.~Primulando, J.~Julio and P.~Uttayarat,
``Scalar phenomenology in type-II seesaw model,''
JHEP \textbf{08}, 024 (2019)
doi:10.1007/JHEP08(2019)024
[arXiv:1903.02493 [hep-ph]].
%32 citations counted in INSPIRE as of 17 Oct 2022

%\cite{Zhou:2022mlz}
\bibitem{Zhou:2022mlz}
R.~Zhou, L.~Bian and Y.~Du,
``Electroweak phase transition and gravitational waves in the type-II seesaw model,''
JHEP \textbf{08}, 205 (2022)
doi:10.1007/JHEP08(2022)205
[arXiv:2203.01561 [hep-ph]].
%5 citations counted in INSPIRE as of 17 Oct 2022

%%%%%%%%%%% ScannerS %%%%%%
%\cite{Muhlleitner:2020wwk}
% \bibitem{Muhlleitner:2020wwk}
% M.~M\"uhlleitner, M.~O.~P.~Sampaio, R.~Santos and J.~Wittbrodt,
% %``ScannerS: parameter scans in extended scalar sectors,''
% Eur. Phys. J. C \textbf{82}, no.3, 198 (2022)
% doi:10.1140/epjc/s10052-022-10139-w
% [arXiv:2007.02985 [hep-ph]].
%51 citations counted in INSPIRE as of 09 Feb 2023

%\cite{Haber:2006ue}
% \bibitem{Haber:2006ue}
% H.~E.~Haber and D.~O'Neil,
% ``Basis-independent methods for the two-Higgs-doublet model. II. The Significance of tan$\beta$,''
% Phys. Rev. D \textbf{74}, 015018 (2006)
% [erratum: Phys. Rev. D \textbf{74}, no.5, 059905 (2006)]
% doi:10.1103/PhysRevD.74.015018
% [arXiv:hep-ph/0602242 [hep-ph]].
%187 citations counted in INSPIRE as of 09 Oct 2022

%======================================================================
%\cite{Herrero-Garcia:2019mcy}
% \bibitem{Herrero-Garcia:2019mcy}
% J.~Herrero-Garcia, M.~Nebot, F.~Rajec, M.~White and A.~G.~Williams,
% ``Higgs Quark Flavor Violation: Simplified Models and Status of General Two-Higgs-Doublet Model,''
% JHEP \textbf{02}, 147 (2020)
% doi:10.1007/JHEP02(2020)147
% [arXiv:1907.05900 [hep-ph]].
%6 citations counted in INSPIRE as of 03 Oct 2022

%======================================================================
%\cite{Athron:2021auq}
% \bibitem{Athron:2021auq}
% P.~Athron, C.~Balazs, T.~E.~Gonzalo, D.~Jacob, F.~Mahmoudi and C.~Sierra,
% ``Likelihood analysis of the flavour anomalies and g \textendash{} 2 in the general two Higgs doublet model,''
% JHEP \textbf{01}, 037 (2022)
% doi:10.1007/JHEP01(2022)037
% [arXiv:2111.10464 [hep-ph]].
%10 citations counted in INSPIRE as of 03 Oct 2022

%\cite{Peskin:1991sw}
% \bibitem{Peskin:1991sw}
% M.~E.~Peskin and T.~Takeuchi,
% ``Estimation of oblique electroweak corrections,''
% Phys. Rev. D \textbf{46}, 381-409 (1992)
% doi:10.1103/PhysRevD.46.381
%2569 citations counted in INSPIRE as of 09 Feb 2023

%\cite{Haber:2010bw}
% \bibitem{Haber:2010bw}
% H.~E.~Haber and D.~O'Neil,
% ``Basis-independent methods for the two-Higgs-doublet model III: The CP-conserving limit, custodial symmetry, and the oblique parameters S, T, U,''
% Phys. Rev. D \textbf{83}, 055017 (2011)
% doi:10.1103/PhysRevD.83.055017
% [arXiv:1011.6188 [hep-ph]].
%208 citations counted in INSPIRE as of 20 Mar 2023

%\cite{Bechtle:2020pkv}
% \bibitem{Bechtle:2020pkv}
% P.~Bechtle, D.~Dercks, S.~Heinemeyer, T.~Klingl, T.~Stefaniak, G.~Weiglein and J.~Wittbrodt,
% ``HiggsBounds-5: Testing Higgs Sectors in the LHC 13 TeV Era,''
% Eur. Phys. J. C \textbf{80}, no.12, 1211 (2020)
% doi:10.1140/epjc/s10052-020-08557-9
% [arXiv:2006.06007 [hep-ph]].
%157 citations counted in INSPIRE as of 09 Feb 2023

%\cite{Bechtle:2020uwn}
% \bibitem{Bechtle:2020uwn}
% P.~Bechtle, S.~Heinemeyer, T.~Klingl, T.~Stefaniak, G.~Weiglein and J.~Wittbrodt,
% ``HiggsSignals-2: Probing new physics with precision Higgs measurements in the LHC 13 TeV era,''
% Eur. Phys. J. C \textbf{81}, no.2, 145 (2021)
% doi:10.1140/epjc/s10052-021-08942-y
% [arXiv:2012.09197 [hep-ph]].
%122 citations counted in INSPIRE as of 09 Feb 2023

%======================================================================
%\cite{WahabElKaffas:2007xd}
% \bibitem{WahabElKaffas:2007xd}
% A.~Wahab El Kaffas, P.~Osland and O.~M.~Ogreid,
% ``Constraining the Two-Higgs-Doublet-Model parameter space,''
% Phys. Rev. D \textbf{76}, 095001 (2007)
% doi:10.1103/PhysRevD.76.095001
% [arXiv:0706.2997 [hep-ph]].
%135 citations counted in INSPIRE as of 04 Oct 2022

%======================================================================
%\cite{Fontes:2017zfn}
% \bibitem{Fontes:2017zfn}
% D.~Fontes, M.~M\"uhlleitner, J.~C.~Rom\~ao, R.~Santos, J.~P.~Silva and J.~Wittbrodt,
% ``The C2HDM revisited,''
% JHEP \textbf{02}, 073 (2018)
% doi:10.1007/JHEP02(2018)073
% [arXiv:1711.09419 [hep-ph]].
%56 citations counted in INSPIRE as of 04 Oct 2022

%======================================================================
%\cite{Khater:2003wq}
% \bibitem{Khater:2003wq}
% W.~Khater and P.~Osland,
% ``CP violation in top quark production at the LHC and two Higgs doublet models,''
% Nucl. Phys. B \textbf{661}, 209-234 (2003)
% doi:10.1016/S0550-3213(03)00300-6
% [arXiv:hep-ph/0302004 [hep-ph]].
%95 citations counted in INSPIRE as of 04 Oct 2022

%======================================================================
%\cite{Inoue:2014nva}
% \bibitem{Inoue:2014nva}
% S.~Inoue, M.~J.~Ramsey-Musolf and Y.~Zhang,
% ``CP-violating phenomenology of flavor conserving two Higgs doublet models,''
% Phys. Rev. D \textbf{89}, no.11, 115023 (2014)
% doi:10.1103/PhysRevD.89.115023
% [arXiv:1403.4257 [hep-ph]].
%136 citations counted in INSPIRE as of 04 Oct 2022

%======================================================================
%\cite{ElKaffas:2007rq}
% \bibitem{ElKaffas:2007rq}
% A.~W.~El Kaffas, P.~Osland and O.~M.~Ogreid,
% ``CP violation, stability and unitarity of the two Higgs doublet model,''
% Nonlin. Phenom. Complex Syst. \textbf{10}, 347-357 (2007)
% [arXiv:hep-ph/0702097 [hep-ph]].
%66 citations counted in INSPIRE as of 04 Oct 2022 

%\cite{Muong-2:2021ojo}
\bibitem{Muong-2:2021ojo}
B.~Abi \textit{et al.} [Muon g-2],
``Measurement of the Positive Muon Anomalous Magnetic Moment to 0.46 ppm,''
Phys. Rev. Lett. \textbf{126}, no.14, 141801 (2021)
doi:10.1103/PhysRevLett.126.141801
[arXiv:2104.03281 [hep-ex]].
%984 citations counted in INSPIRE as of 05 Oct 2022

	%\cite{Muong-2:2006rrc}
\bibitem{Muong-2:2006rrc}
G.~W.~Bennett \textit{et al.} [Muon g-2 Collaboration],
%``Final Report of the Muon E821 Anomalous Magnetic Moment Measurement at BNL,''
Phys. Rev. D \textbf{73}, 072003 (2006)
doi:10.1103/PhysRevD.73.072003
[arXiv:hep-ex/0602035 [hep-ex]].
%2918 citations counted in INSPIRE as of 20 Jul 2022

%----------  Muon g-2  ---------
%\cite{Aoyama:2012wk}
\bibitem{Aoyama:2012wk}
T.~Aoyama, M.~Hayakawa, T.~Kinoshita and M.~Nio,
%``Complete Tenth-Order QED Contribution to the Muon g-2,''
Phys. Rev. Lett. \textbf{109}, 111808 (2012)
doi:10.1103/PhysRevLett.109.111808
[arXiv:1205.5370 [hep-ph]].
%549 citations counted in INSPIRE as of 01 Sep 2022

%\cite{Aoyama:2019ryr}
\bibitem{Aoyama:2019ryr}
T.~Aoyama, T.~Kinoshita and M.~Nio,
%``Theory of the Anomalous Magnetic Moment of the Electron,''
Atoms \textbf{7}, no.1, 28 (2019)
doi:10.3390/atoms7010028
%214 citations counted in INSPIRE as of 26 Aug 2022

%\cite{Czarnecki:2002nt}
\bibitem{Czarnecki:2002nt}
A.~Czarnecki, W.~J.~Marciano and A.~Vainshtein,
%``Refinements in electroweak contributions to the muon anomalous magnetic moment,''
Phys. Rev. D \textbf{67}, 073006 (2003)
[erratum: Phys. Rev. D \textbf{73}, 119901 (2006)]
doi:10.1103/PhysRevD.67.073006
[arXiv:hep-ph/0212229 [hep-ph]].
%496 citations counted in INSPIRE as of 01 Sep 2022

%\cite{Gnendiger:2013pva}
\bibitem{Gnendiger:2013pva}
C.~Gnendiger, D.~St\"ockinger and H.~St\"ockinger-Kim,
%``The electroweak contributions to $(g-2)_\mu$ after the Higgs boson mass measurement,''
Phys. Rev. D \textbf{88}, 053005 (2013)
doi:10.1103/PhysRevD.88.053005
[arXiv:1306.5546 [hep-ph]].
%386 citations counted in INSPIRE as of 01 Sep 2022

%\cite{Davier:2017zfy}
\bibitem{Davier:2017zfy}
M.~Davier, A.~Hoecker, B.~Malaescu and Z.~Zhang,
%``Reevaluation of the hadronic vacuum polarisation contributions to the Standard Model predictions of the muon $g-2$ and ${\alpha (m_Z^2)}$ using newest hadronic cross-section data,''
Eur. Phys. J. C \textbf{77}, no.12, 827 (2017)
doi:10.1140/epjc/s10052-017-5161-6
[arXiv:1706.09436 [hep-ph]].
%464 citations counted in INSPIRE as of 01 Sep 2022

%\cite{Keshavarzi:2018mgv}
\bibitem{Keshavarzi:2018mgv}
A.~Keshavarzi, D.~Nomura and T.~Teubner,
%``Muon $g-2$ and $\alpha(M_Z^2)$: a new data-based analysis,''
Phys. Rev. D \textbf{97}, no.11, 114025 (2018)
doi:10.1103/PhysRevD.97.114025
[arXiv:1802.02995 [hep-ph]].
%543 citations counted in INSPIRE as of 01 Sep 2022

%\cite{Colangelo:2018mtw}
\bibitem{Colangelo:2018mtw}
G.~Colangelo, M.~Hoferichter and P.~Stoffer,
%``Two-pion contribution to hadronic vacuum polarization,''
JHEP \textbf{02}, 006 (2019)
doi:10.1007/JHEP02(2019)006
[arXiv:1810.00007 [hep-ph]].
%269 citations counted in INSPIRE as of 01 Sep 2022

%\cite{Hoferichter:2019mqg}
\bibitem{Hoferichter:2019mqg}
M.~Hoferichter, B.~L.~Hoid and B.~Kubis,
%``Three-pion contribution to hadronic vacuum polarization,''
JHEP \textbf{08}, 137 (2019)
doi:10.1007/JHEP08(2019)137
[arXiv:1907.01556 [hep-ph]].
%218 citations counted in INSPIRE as of 01 Sep 2022

%\cite{Davier:2019can}
\bibitem{Davier:2019can}
M.~Davier, A.~Hoecker, B.~Malaescu and Z.~Zhang,
%``A new evaluation of the hadronic vacuum polarisation contributions to the muon anomalous magnetic moment and to $\mathbf{\boldsymbol\alpha(m_Z^2)}$,''
Eur. Phys. J. C \textbf{80}, no.3, 241 (2020)
[erratum: Eur. Phys. J. C \textbf{80}, no.5, 410 (2020)]
doi:10.1140/epjc/s10052-020-7792-2
[arXiv:1908.00921 [hep-ph]].
%416 citations counted in INSPIRE as of 01 Sep 2022

%\cite{Keshavarzi:2019abf}
\bibitem{Keshavarzi:2019abf}
A.~Keshavarzi, D.~Nomura and T.~Teubner,
%``$g-2$ of charged leptons, $\alpha (M^2_Z)$ , and the hyperfine splitting of muonium,''
Phys. Rev. D \textbf{101}, no.1, 014029 (2020)
doi:10.1103/PhysRevD.101.014029
[arXiv:1911.00367 [hep-ph]].
%305 citations counted in INSPIRE as of 01 Sep 2022

%\cite{Kurz:2014wya}
\bibitem{Kurz:2014wya}
A.~Kurz, T.~Liu, P.~Marquard and M.~Steinhauser,
%``Hadronic contribution to the muon anomalous magnetic moment to next-to-next-to-leading order,''
Phys. Lett. B \textbf{734}, 144-147 (2014)
doi:10.1016/j.physletb.2014.05.043
[arXiv:1403.6400 [hep-ph]].
%338 citations counted in INSPIRE as of 01 Sep 2022

%\cite{Melnikov:2003xd}
\bibitem{Melnikov:2003xd}
K.~Melnikov and A.~Vainshtein,
%``Hadronic light-by-light scattering contribution to the muon anomalous magnetic moment revisited,''
Phys. Rev. D \textbf{70}, 113006 (2004)
doi:10.1103/PhysRevD.70.113006
[arXiv:hep-ph/0312226 [hep-ph]].
%513 citations counted in INSPIRE as of 01 Sep 2022

%\cite{Masjuan:2017tvw}
\bibitem{Masjuan:2017tvw}
P.~Masjuan and P.~Sanchez-Puertas,
%``Pseudoscalar-pole contribution to the $(g_{\mu}-2)$: a rational approach,''
Phys. Rev. D \textbf{95}, no.5, 054026 (2017)
doi:10.1103/PhysRevD.95.054026
[arXiv:1701.05829 [hep-ph]].
%217 citations counted in INSPIRE as of 01 Sep 2022

%\cite{Colangelo:2017fiz}
\bibitem{Colangelo:2017fiz}
G.~Colangelo, M.~Hoferichter, M.~Procura and P.~Stoffer,
%``Dispersion relation for hadronic light-by-light scattering: two-pion contributions,''
JHEP \textbf{04}, 161 (2017)
doi:10.1007/JHEP04(2017)161
[arXiv:1702.07347 [hep-ph]].
%273 citations counted in INSPIRE as of 01 Sep 2022

%\cite{Hoferichter:2018kwz}
\bibitem{Hoferichter:2018kwz}
M.~Hoferichter, B.~L.~Hoid, B.~Kubis, S.~Leupold and S.~P.~Schneider,
%``Dispersion relation for hadronic light-by-light scattering: pion pole,''
JHEP \textbf{10}, 141 (2018)
doi:10.1007/JHEP10(2018)141
[arXiv:1808.04823 [hep-ph]].
%245 citations counted in INSPIRE as of 01 Sep 2022

%\cite{Gerardin:2019vio}
\bibitem{Gerardin:2019vio}
A.~G\'erardin, H.~B.~Meyer and A.~Nyffeler,
%``Lattice calculation of the pion transition form factor with $N_f=2+1$ Wilson quarks,''
Phys. Rev. D \textbf{100}, no.3, 034520 (2019)
doi:10.1103/PhysRevD.100.034520
[arXiv:1903.09471 [hep-lat]].
%207 citations counted in INSPIRE as of 01 Sep 2022

%\cite{Bijnens:2019ghy}
\bibitem{Bijnens:2019ghy}
J.~Bijnens, N.~Hermansson-Truedsson and A.~Rodr\'\i{}guez-S\'anchez,
%``Short-distance constraints for the HLbL contribution to the muon anomalous magnetic moment,''
Phys. Lett. B \textbf{798}, 134994 (2019)
doi:10.1016/j.physletb.2019.134994
[arXiv:1908.03331 [hep-ph]].
%184 citations counted in INSPIRE as of 01 Sep 2022

%\cite{Colangelo:2019uex}
\bibitem{Colangelo:2019uex}
G.~Colangelo, F.~Hagelstein, M.~Hoferichter, L.~Laub and P.~Stoffer,
%``Longitudinal short-distance constraints for the hadronic light-by-light contribution to $(g-2)_\mu$ with large-$N_c$ Regge models,''
JHEP \textbf{03}, 101 (2020)
doi:10.1007/JHEP03(2020)101
[arXiv:1910.13432 [hep-ph]].
%192 citations counted in INSPIRE as of 01 Sep 2022

%\cite{Blum:2019ugy}
\bibitem{Blum:2019ugy}
T.~Blum, N.~Christ, M.~Hayakawa, T.~Izubuchi, L.~Jin, C.~Jung and C.~Lehner,
%``Hadronic Light-by-Light Scattering Contribution to the Muon Anomalous Magnetic Moment from Lattice QCD,''
Phys. Rev. Lett. \textbf{124}, no.13, 132002 (2020)
doi:10.1103/PhysRevLett.124.132002
[arXiv:1911.08123 [hep-lat]].
%215 citations counted in INSPIRE as of 01 Sep 2022

%\cite{Colangelo:2014qya}
\bibitem{Colangelo:2014qya}
G.~Colangelo, M.~Hoferichter, A.~Nyffeler, M.~Passera and P.~Stoffer,
%``Remarks on higher-order hadronic corrections to the muon g\ensuremath{-}2,''
Phys. Lett. B \textbf{735}, 90-91 (2014)
doi:10.1016/j.physletb.2014.06.012
[arXiv:1403.7512 [hep-ph]].
%287 citations counted in INSPIRE as of 01 Sep 2022


%%%%%% g-2 SM review  %%%%%%

%\cite{Aoyama:2020ynm}
\bibitem{Aoyama:2020ynm}
T.~Aoyama, \textit{et al.}
%``The anomalous magnetic moment of the muon in the Standard Model,''
Phys. Rept. \textbf{887}, 1-166 (2020)
doi:10.1016/j.physrep.2020.07.006
[arXiv:2006.04822 [hep-ph]].
%321 citations counted in INSPIRE as of 09 Sep 2021
%=====================

%\cite{Cheung:2001hz}
\bibitem{Cheung:2001hz}
K.~m.~Cheung, C.~H.~Chou and O.~C.~W.~Kong,
``Muon anomalous magnetic moment, two Higgs doublet model, and supersymmetry,''
Phys. Rev. D \textbf{64}, 111301 (2001)
doi:10.1103/PhysRevD.64.111301
[arXiv:hep-ph/0103183 [hep-ph]].
%134 citations counted in INSPIRE as of 05 Oct 2022

%\cite{Cheung:2003pw}
\bibitem{Cheung:2003pw}
K.~Cheung and O.~C.~W.~Kong,
``Can the two Higgs doublet model survive the constraint from the muon anomalous magnetic moment as suggested?,''
Phys. Rev. D \textbf{68}, 053003 (2003)
doi:10.1103/PhysRevD.68.053003
[arXiv:hep-ph/0302111 [hep-ph]].
%87 citations counted in INSPIRE as of 05 Oct 2022

%\cite{Zhou:2001ew}
\bibitem{Zhou:2001ew}
Y.~F.~Zhou and Y.~L.~Wu,
``Lepton flavor changing scalar interactions and muon g-2,''
Eur. Phys. J. C \textbf{27}, 577-585 (2003)
doi:10.1140/epjc/s2003-01137-1
[arXiv:hep-ph/0110302 [hep-ph]].
%19 citations counted in INSPIRE as of 05 Oct 2022

%\cite{Aoki:2009ha}
\bibitem{Aoki:2009ha}
M.~Aoki, S.~Kanemura, K.~Tsumura and K.~Yagyu,
``Models of Yukawa interaction in the two Higgs doublet model, and their collider phenomenology,''
Phys. Rev. D \textbf{80}, 015017 (2009)
doi:10.1103/PhysRevD.80.015017
[arXiv:0902.4665 [hep-ph]].
%338 citations counted in INSPIRE as of 18 Oct 2022

%\cite{Cao:2009as}
\bibitem{Cao:2009as}
J.~Cao, P.~Wan, L.~Wu and J.~M.~Yang,
``Lepton-Specific Two-Higgs Doublet Model: Experimental Constraints and Implication on Higgs Phenomenology,''
Phys. Rev. D \textbf{80}, 071701 (2009)
doi:10.1103/PhysRevD.80.071701
[arXiv:0909.5148 [hep-ph]].
%74 citations counted in INSPIRE as of 09 Feb 2023

%\cite{Han:2022juu}
\bibitem{Han:2022juu}
X.~F.~Han, F.~Wang, L.~Wang, J.~M.~Yang and Y.~Zhang,
``Joint explanation of W-mass and muon g\textendash{}2 in the 2HDM*,''
Chin. Phys. C \textbf{46}, no.10, 103105 (2022)
doi:10.1088/1674-1137/ac7c63
[arXiv:2204.06505 [hep-ph]].
%56 citations counted in INSPIRE as of 05 Oct 2022

%\cite{Ferreira:2021gke}
\bibitem{Ferreira:2021gke}
P.~M.~Ferreira, B.~L.~Gon\c{c}alves, F.~R.~Joaquim and M.~Sher,
``(g-2)\ensuremath{\mu} in the 2HDM and slightly beyond: An updated view,''
Phys. Rev. D \textbf{104}, no.5, 053008 (2021)
doi:10.1103/PhysRevD.104.053008
[arXiv:2104.03367 [hep-ph]].
%29 citations counted in INSPIRE as of 05 Oct 2022

%\cite{Kim:2022xuo}
\bibitem{Kim:2022xuo}
J.~Kim,
``Compatibility of muon g-2, W mass anomaly in type-X 2HDM,''
Phys. Lett. B \textbf{832}, 137220 (2022)
doi:10.1016/j.physletb.2022.137220
[arXiv:2205.01437 [hep-ph]].
%14 citations counted in INSPIRE as of 05 Oct 2022

%\cite{Pich:2009sp}
\bibitem{Pich:2009sp}
A.~Pich and P.~Tuzon,
``Yukawa Alignment in the Two-Higgs-Doublet Model,''
Phys. Rev. D \textbf{80}, 091702 (2009)
doi:10.1103/PhysRevD.80.091702
[arXiv:0908.1554 [hep-ph]].
%342 citations counted in INSPIRE as of 14 Oct 2022

%======================================================================
%\cite{Eberhardt:2020dat}
\bibitem{Eberhardt:2020dat}
O.~Eberhardt, A.~P.~Mart\'\i{}nez and A.~Pich,
``Global fits in the Aligned Two-Higgs-Doublet model,''
JHEP \textbf{05}, 005 (2021)
doi:10.1007/JHEP05(2021)005
[arXiv:2012.09200 [hep-ph]].
%29 citations counted in INSPIRE as of 03 Oct 2022

%\cite{Ilisie:2015tra}
\bibitem{Ilisie:2015tra}
V.~Ilisie,
``New Barr-Zee contributions to $\mathbf{(g-2)_\mu}$ in two-Higgs-doublet models,''
JHEP \textbf{04}, 077 (2015)
doi:10.1007/JHEP04(2015)077
[arXiv:1502.04199 [hep-ph]].
%104 citations counted in INSPIRE as of 05 Oct 2022

%======================================================================
%\cite{Davidson:2005cw}
\bibitem{Davidson:2005cw}
S.~Davidson and H.~E.~Haber,
``Basis-independent methods for the two-Higgs-doublet model,''
Phys. Rev. D \textbf{72}, 035004 (2005)
[erratum: Phys. Rev. D \textbf{72}, 099902 (2005)]
doi:10.1103/PhysRevD.72.099902
[arXiv:hep-ph/0504050 [hep-ph]].
%424 citations counted in INSPIRE as of 09 Oct 2022


\end{thebibliography}
\end{document}